\documentclass[acmsmall]{acmart}

\AtBeginDocument{%
  }

\usepackage{soul}
\usepackage{url}
\usepackage{epstopdf}
\usepackage{float}
\usepackage[utf8]{inputenc}
\usepackage{graphicx}
\usepackage{amsmath}
\usepackage{algorithm}
\usepackage[noend]{algorithmic}
\usepackage{comment}
\usepackage{pifont}
\usepackage{smartdiagram}
\usepackage{diagbox}
\usepackage{mathtools}
\usepackage{amsthm}
\usepackage{bm}
\usepackage{lscape}
\usepackage{tikz}
\usepackage{pgfplots}
\usepackage{listings}
\usepackage{wrapfig}
\usepackage{xr}

\usepackage{graphicx,array}
\usepackage{color,soul,xspace}
\usepackage{comment}
\usepackage{epsfig}
\usepackage{subfig}
\usepackage{caption}
\usepackage[figuresright]{rotating}
\usepackage{booktabs}

\usepackage{color, soul}
\usepackage{xcolor}

\usepackage{pifont}
\usepackage{array}

\usepackage[flushleft]{threeparttable}
\usepackage{tablefootnote}
\usepackage{pdfrender}
\newcommand{\dashcheckmark}{
    \textpdfrender{
        TextRenderingMode=Stroke,
        LineWidth=0.5pt,
        LineDashPattern=[1 1]0,
    }{\checkmark}
}

\newcommand{\cmark}{\ding{51}}%

\newcommand{\sorec}{SocialRS\xspace}

\newcommand{\ie}{\textit i.e.,\xspace}
\newcommand{\eg}{\textit e.g.,\xspace}

\newtheorem{prob}{\textbf{Problem}}

\usepackage{multirow}
\usepackage{multicol}

\setcopyright{acmcopyright}
\copyrightyear{2024}
\acmYear{2024}
\acmDOI{10.1145/3661821}

\acmJournal{CSUR}
\acmMonth{4}

\begin{document}

\title{A Survey of Graph Neural Networks for Social Recommender Systems}

\author{Kartik Sharma$^*$}
\thanks{$^*$ Equal Contribution}
\email{ksartik@gatech.edu}
\affiliation{
    \institution{Georgia Institute of Technology}
    \country{USA}
}

\author{Yeon-Chang Lee$^*$} 
\email{yeonchang@unist.ac.kr}
\affiliation{
    \institution{UNIST}
    \country{South Korea}
}

\author{Sivagami Nambi}
\email{sivagami.nambi@gatech.edu}
\affiliation{
    \institution{Georgia Institute of Technology}
    \country{USA}
}

\author{Aditya Salian$^\dagger$}
\thanks{$^\dagger$ Equal Contribution}
\email{asalian@gatech.edu}
\affiliation{
    \institution{Georgia Institute of Technology}
    \country{USA}
}

\author{Shlok Shah$^\dagger$}
\email{sshah672@gatech.edu}
\affiliation{
    \institution{Georgia Institute of Technology}
    \country{USA}
}

\author{Sang-Wook Kim}
\email{wook@hanyang.ac.kr}
\affiliation{
    \institution{Hanyang University}
    \country{South Korea}
}

\author{Srijan Kumar$^+$}
\thanks{$^+ $ Corresponding author}
\email{srijan@gatech.edu}
\affiliation{
    \institution{Georgia Institute of Technology}
    \country{USA}
}

\renewcommand{\shortauthors}{Sharma \textit{et al.}}

\begin{abstract}
Social recommender systems (\sorec) simultaneously leverage the user-to-item interactions as well as the user-to-user social relations for the task of generating item recommendations to users.
Additionally exploiting social relations is clearly effective in understanding users' tastes due to the effects of homophily and social influence.
For this reason, \sorec has increasingly attracted attention.
In particular, with the advance of graph neural networks (GNN), many GNN-based \sorec methods have been developed recently.
Therefore, we conduct a comprehensive and systematic review of the literature on GNN-based \sorec. 

In this survey, we first identify $84$ papers on GNN-based \sorec after annotating 2,151 papers by following the PRISMA framework (preferred reporting items for systematic reviews and meta-analyses).
Then, we comprehensively review them in terms of their inputs and architectures to propose a novel taxonomy: (1) input taxonomy includes 5 groups of input type notations and 7 groups of input representation notations; (2) architecture taxonomy includes 8 groups of GNN encoder notations, 2 groups of decoder notations, and 12 groups of loss function notations. 
We classify the GNN-based \sorec methods into several categories as per the taxonomy and describe their details.
Furthermore, we summarize benchmark datasets and metrics widely used to evaluate the GNN-based \sorec methods.
Finally, we conclude this survey by presenting some future research directions.
GitHub repository with the curated list of papers are available at \url{https://github.com/claws-lab/awesome-GNN-social-recsys}.

\end{abstract}



\maketitle

\section{Introduction}

With the advent of online social network platforms (\eg Facebook, Twitter, Instagram, etc.), there has been a surge of research efforts in developing social recommender systems (\sorec), which simultaneously utilize user-user social relations along with user-item interactions to recommend relevant items to users.
Exploiting social relations in recommendation works well because of the effects of \textit{social homophily}~\cite{mcpherson2001birds} and \textit{social influence}~\cite{marsden1993network}:
(1) social homophily indicates that a user tends to connect herself to other users with similar attributes and preferences, and
(2) social influence indicates that users with direct or indirect relations tend to influence each other to make themselves become more similar.
Accordingly, \sorec can effectively mitigate the data sparsity problem by exploiting social neighbors to capture the preferences of a sparsely interacting user.

Literature has shown that \sorec can be applied successfully in various recommendation domains (\eg product~\cite{wu2020diffnet++, wu2019danser}, music~\cite{yu2021sept,yu2021mhcn,yu2022esrf}, location~\cite{wu2022dcrec,seng2021atgcn,li2022atstggnn}, and image~\cite{wu2019diffnet,wu2022eagcn,tao2022design}), thereby improving user satisfaction. 
Furthermore, techniques and insights explored from \sorec can also be exploited in real-world applications other than recommendations.
For instance, García-Sánchez et al.~\cite{GarciaSanchezP20} leveraged \sorec to design a decision-making system for marketing (\eg advertisement), while Gasparetti et al.~\cite{GasparettiSM21} analyzed \sorec in terms of community detection. 

\usetikzlibrary{patterns}
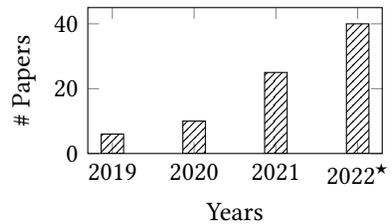
\begin{wrapfigure}{r}{5.5cm} 
\centering
\begin{tikzpicture}
    \begin{axis}[
        width=5.5cm,
        height=3.5cm,
        bar width=0.3cm,
        xlabel={Years},
        ylabel={\# Papers},
        ymin=0, ymax=45,
        xticklabels={2019, 2020, 2021, 2022$^\star$},
        xtick={1, 2, 3, 4},
        ylabel near ticks,
        xlabel near ticks,
        ]
        \addplot[ybar,fill=black, pattern color = black, pattern=north east lines] coordinates {
        (1,6)(2,10)(3,25)(4,40)        };
    \end{axis}
    \end{tikzpicture}
    \vspace{-0.3cm}
    \caption{The number of papers related to GNN-based \sorec per year. $^\star$ For 2022, we count the number of relevant papers published until October.}
    \label{fig:pub}
\end{wrapfigure}

Motivated by such wide applicability, there has been an increasing interest in research on developing accurate \sorec models.
In the early days, research focused on matrix factorization (MF) techniques~\cite{MaYLK08a,MaKL09,MaLK09,JamaliE10,MaZLLK11,0002LLL13,TangHGL13}.
However, MF-based methods cannot effectively model the complex (\ie non-linear) relationships inherent in user-user social relations and user-item interactions~\cite{shokeen2020social}.
Motivated by this, most recent works have focused on applying deep-learning techniques to \sorec, \eg autoencoder~\cite{DengHXWW17,YingCXW16}, generative adversarial networks (GAN)~\cite{krishnan2019modular}, and graph neural networks (GNN)~\cite{fan2019graphrec,wu2019diffnet}.

In particular, since user-item interactions and user-user social relations can naturally be represented as graph data, GNN-based \sorec has increasingly attracted attention in the literature.
As a demonstration, Figure~\ref{fig:pub} shows that the number of papers related to GNN-based \sorec has increased consistently since 2019.
Given the growing and timely interest in this area, we survey GNN-based \sorec methods in this survey.

\subsection{Challenges}

Applying GNN into \sorec is not trivial and faces the following challenges.

\vspace{1mm}
\textbf{Input representation}. The input data should be modeled appropriately into a heterogeneous graph structure. Many \sorec methods build two separate graphs: one where nodes represent users and items, and edges represent user-item interactions; the other where nodes represent users and edges represent user-user social relations. 
Thus, GNN methods for \sorec need to extract knowledge from both the networks simultaneously for accurate inference. This is in contrast with most regular GNNs that consider only a single network.
Additionally, we note that there are valuable input features in the two networks, such as user/item attributes, item knowledge/relation, and group information. Thus, methods fuse features along with network information in GNN-based \sorec.
In this survey, we discuss the input types used in GNN-based \sorec methods and the different ways they are represented as graphs.

\vspace{1mm}
\textbf{Design of GNN encoder}. The performance of GNN-based \sorec methods relies heavily on their GNN encoders, which aim to represent users and items into low-dimensional embeddings. 
For this reason, existing \sorec methods have explored various design choices regarding GNN encoders and have adopted different architectures according to their goals. 
For instance, many \sorec methods employ the graph attention neural network (GANN)~\cite{velivckovic2017graph} to differentiate each user's preference for items or each user's influence on their social friends.
On the other hand, some methods~\cite{sun2020dgarecr,yan2022ssdrec,gu2021egfrec,narang2021fuserec,niu2021mgsr} use the graph recurrent neural networks (GRNN)~\cite{peng2017cross,zayats2018conversation} to model the sequential behaviors of users.
It should be noted that GNN encoders for \sorec need to simultaneously consider the characteristics of user-item interactions and user-user social relations. 
This is in contrast with GNN encoders for non-\sorec that model only user-item interactions.
In this survey, we discuss different types of GNN encoders used by \sorec methods.

\vspace{1mm}
\textbf{Training}. The training of GNN-based \sorec should be designed to reflect users' tastes and items' characteristics in the embeddings for the corresponding users and items. To this end, \sorec methods employ well-known loss functions, such as mean squared error (MSE), Bayesian personalized ranking (BPR)~\cite{RendleFGS09}, and cross-entropy (CE), to reconstruct user behaviors. Furthermore, to mitigate the data sparsity problem, some works have additionally employed auxiliary loss functions such as self-supervised loss~\cite{liu2020selfsupervised} and group-based loss~\cite{leng2022glow,liao2022gman}.
It is worth mentioning that loss functions used by GNN-based \sorec are designed so that rich structural information such as motifs and user attributes can be exploited. These are not considered by loss functions for non-\sorec.
In this survey, we discuss the training remedies of GNN-based \sorec methods to learn the user and item embeddings.

\subsection{Related Surveys}

\begin{table*}[t]
    \centering
    \begin{threeparttable}
    \caption{Comparison with existing surveys. For each survey, we summarize the topics covered, some statistics regarding GNN-based \sorec papers (\ie relevant papers), and the main scope to survey.}\label{tab:salesman}
    \newcolumntype{H}{>{\setbox0=\hbox\bgroup}c<{\egroup}@{}}
    \begin{tabular}{c|cHc|cc|c}
        \toprule
        \multirow{2}{*}{\textbf{Surveys}} & \multicolumn{3}{c|}{\textbf{Topics}} & 
        \multicolumn{2}{c|}{\textbf{GNN-based \sorec Papers}} & 
        \multirow{2}{*}{\textbf{Scope}} \\
        & \textbf{\color{red}\sorec} & & \textbf{\color{blue}GNN} & \textbf{\# Papers} & \textbf{Latest Year} &  \\
        \midrule
        \cite{TangHL13,YangGLS14,XuWZC15,DouYD16,ChenHCWZK18} & \cmark & & & $0$ & - & Traditional {\color{red}\sorec} \\
        \cite{GasparettiSM21} & \cmark & & & $0$ 
        & - & {\color{red}\sorec} for CD\\
        \cite{shokeen2020studyFeatures} & \cmark &  &  & $0$ 
        & - & General {\color{red}\sorec} \\
        \cite{shokeen2020social} & \cmark & \dashcheckmark & \dashcheckmark & $1$ 
        & 2019 & General {\color{red}\sorec} \\
        \cite{Deng22} & \dashcheckmark & \dashcheckmark & \dashcheckmark & $2$ 
        & 2019 & Graph-based RS \\
        \cite{wangHW0SOC0Y21} & \dashcheckmark & \dashcheckmark & \dashcheckmark & 3 & 2020 & Graph-based RS \\        
        \cite{wu2020gnnsurvey} & \dashcheckmark & \dashcheckmark & \cmark & $14$ & 
        2021 & {\color{blue}GNN}-based RS \\
        \cite{gao2021gnnsurvey} & \dashcheckmark & \dashcheckmark & \cmark & $19$ & 
        2021 & {\color{blue}GNN}-based RS \\ \midrule
        Ours & \cmark & \cmark & \cmark & $80$ & Oct, 2022 & {\color{blue}GNN}-based {\color{red}\sorec}  \\
        
        \bottomrule
    \end{tabular}
    \begin{tablenotes}
        \footnotesize
        \item \cmark: fully covered, \dashcheckmark: partially covered. 
        \item \# Papers: the number of GNN-based \sorec papers included in the survey. 
        \item Latest year: the latest publication year of a relevant paper included in the survey.
    \end{tablenotes}
    \end{threeparttable}
\end{table*}

Most of the existing surveys, which fully cover \sorec papers, focus either on traditional methods~\cite{PapadimitriouSM12,TangHL13,YangGLS14,XuWZC15,DouYD16,ChenHCWZK18,ShokeenR18} (\eg matrix factorization), feature information~\cite{shokeen2020studyFeatures} (\eg context), or a specific application~\cite{GasparettiSM21} (\eg community detection). 
On the other hand, the other related surveys~\cite{Deng22,wangHW0SOC0Y21,wu2020gnnsurvey,gao2021gnnsurvey} focus on graph-based recommender systems, including GNN-based RS methods, but they partially cover \sorec papers in their surveys.
A comparison between the current survey and the previous surveys is shown in Table~\ref{tab:salesman}.

\begin{figure*}[t]
\centering
  \includegraphics[width=\linewidth]{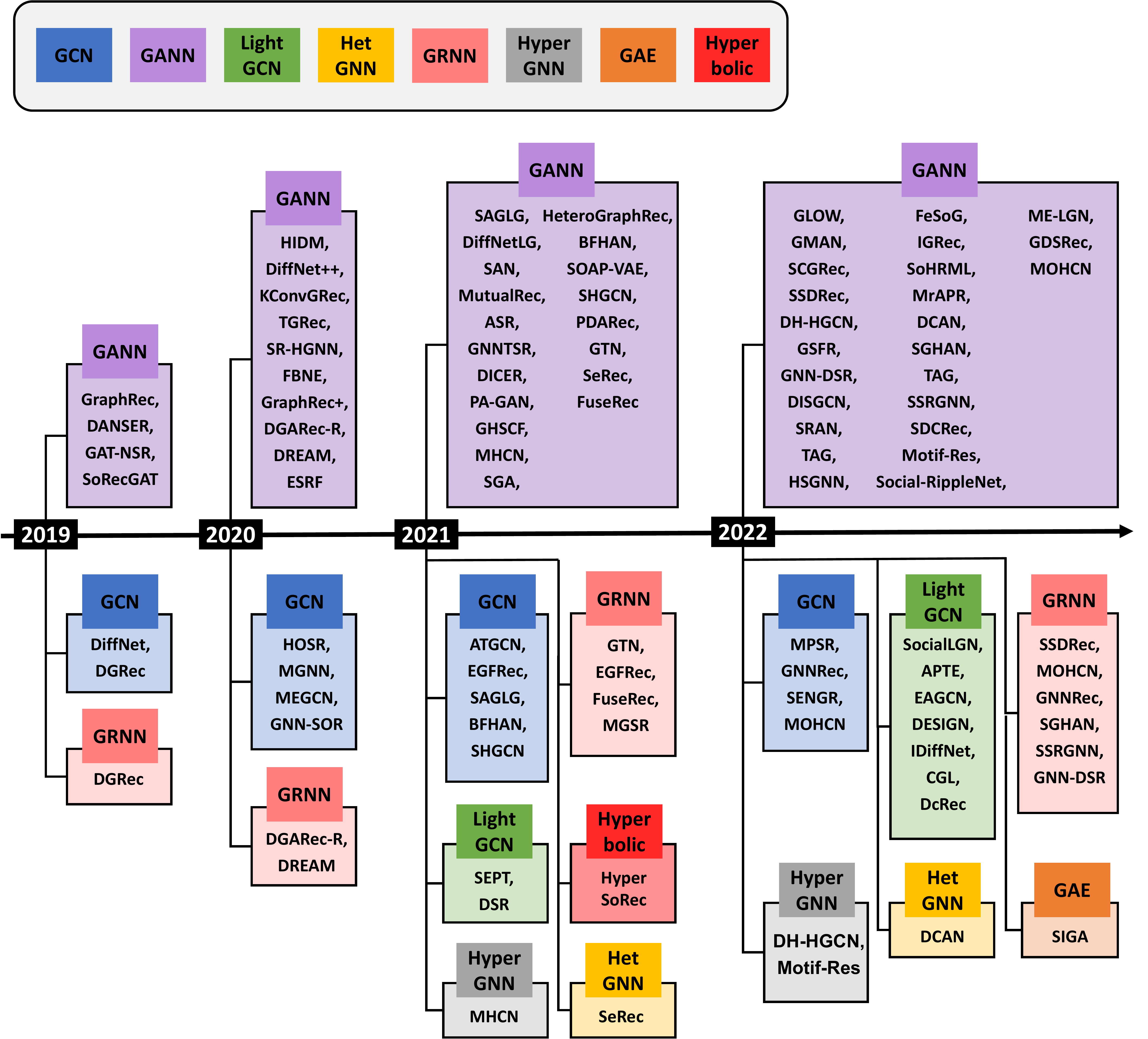}
\caption{A timeline of GNN-based \sorec methods. We categorize methods according to their GNN encoders: graph convolutional network (GCN), lightweight GCN (LightGCN), graph attention neural networks (GANN), heterogeneous GNN (HetGNN), graph recurrent neural networks (GRNN), hypergraph neural networks (HyperGNN), graph autoencoder (GAE), and hyperbolic GNN. It should be noted that some methods employ two or more GNN encoders in their architectures.}
\vspace{-0.5cm}
\label{fig:timeline}
\end{figure*}

Specifically, several survey papers on \sorec have been published before 2019 ~\cite{PapadimitriouSM12,TangHL13,YangGLS14,XuWZC15,DouYD16,ChenHCWZK18,ShokeenR18}. 
However, they only focus on traditional methods such as matrix factorization and collaborative filtering. 
These surveys largely ignore methods that use modern-day deep-learning techniques, in particular GNN. 

More recent surveys discuss the taxonomy of social recommendation, 
starting the comparison of deep-learning based techniques~\cite{shokeen2020studyFeatures,shokeen2020social,GasparettiSM21}. 
However, Shokeen and Rana~\cite{shokeen2020studyFeatures} 
only focus on the taxonomy of feature information regarding social relations,
such as context, trust, and group, used in \sorec methods, while Gasparetti et al.~\cite{GasparettiSM21} only discuss \sorec methods using community detection (CD) techniques. 
Shokeen and Rana~\cite{shokeen2020social} include just one social recommendation method based on GNNs.

With the advent of GNNs in recommender systems, multiple surveys have been conducted on graph-based recommender systems~\cite{wangHW0SOC0Y21,wu2020gnnsurvey,gao2021gnnsurvey,Deng22}. 
However, their focus is not on \sorec as they consider different kinds of recommender systems where graph-learning is employed. They cover only a small section of most representative papers on GNN-based \sorec. Thus, one cannot rely on these surveys to gain insights on the ever-increasing field of using GNNs for \sorec.

As shown in Table~\ref{tab:salesman}, no survey paper exists in the literature that focuses specifically on GNN-based \sorec methods. In the current work, we aim to fill this gap by providing a comprehensive and systematic survey on GNN-based \sorec methods.

\begin{figure}[t]
\centering
\begin{tikzpicture}
    \begin{axis}[
        width=9cm,
        height=11cm,
        bar width=0.2cm,
        xlabel={\# Papers},
        ylabel={Venues},
        xmin=0, xmax=9,
        ymin=0, ymax=22,
        yticklabels={ACM TIST, Data Min. Knowl. Discov., IEEE Intell. Syst., IJCNN, IEEE BigData, IEEE ICDE, IEEE ICDM, ECML-PKDD, ACM KDD, Appl. Intell., Information Sciences, Knowl. Based Syst., IEEE ICTAI, ACM TOIS, DASFAA, ACM WSDM, ACM CIKM, Neurocomputing, ACM SIGIR, WWW, IEEE TKDE},
        ytick={1, 2, 3, 4, 5, 6, 7, 8, 9, 10, 11, 12, 13, 14, 15, 16, 17, 18, 19, 20, 21},
        ylabel near ticks,
        xlabel near ticks,
        ]
        \addplot[xbar,fill=black, pattern color = black, pattern=north east lines] coordinates 
        {
        (1,1)
        (1,2)
        (1,3)
        (1,4)
        (1,5)
        (1,6)
        (1,7)
        (1,8)
        (1,9)
        (2,10)
        (2,11)
        (2,12)
        (2,13)
        (2,14)
        (2,15)
        (2,16)
        (3,17)
        (4,18)
        (4,19)
        (6,20)
        (8,21)
        };
    \end{axis}
    \end{tikzpicture}
    \caption{The number of GNN-based \sorec papers published in relevant journals and conferences. We only present statistics with respect to prominent data mining journals (including IEEE TKDE, ACM TOIS, Knowledge-Based Systems, and Information Sciences) and conferences (including WWW, ACM SIGIR, ACM KDD, ACM CIKM, ACM WSDM, IEEE ICDE, and IEEE ICDM). We believe it would help researchers in this field to identify appropriate venues where GNN-based \sorec papers are published.}\label{fig:venue}
\end{figure}
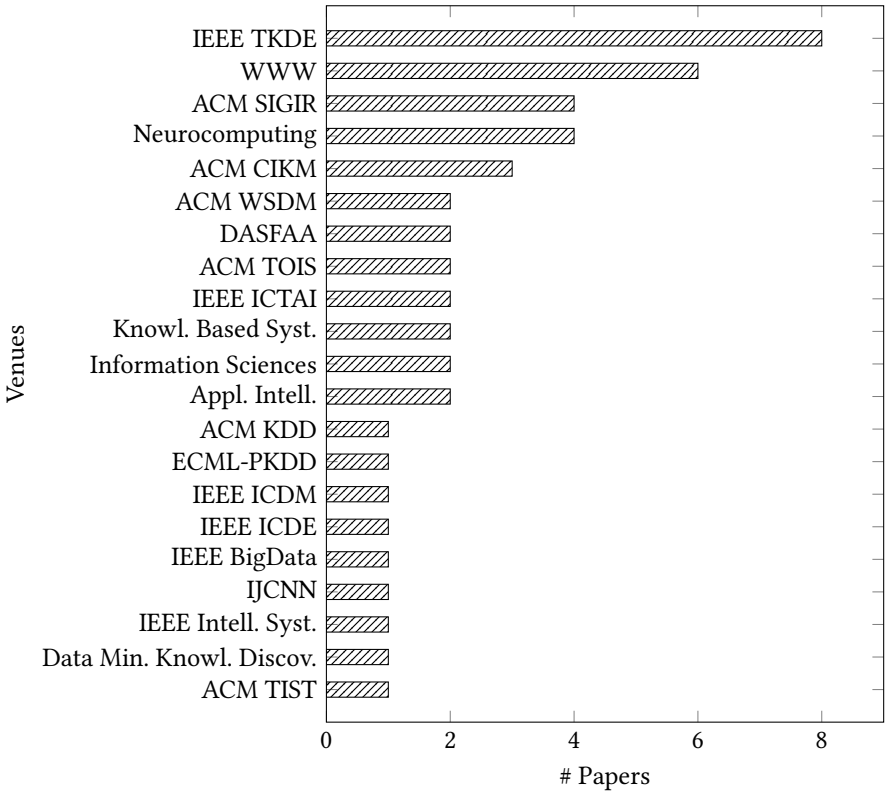

\subsection{Contributions}

The main contribution of this survey paper is summarized as follows:

\begin{itemize}
    \item \textbf{The First Survey in GNN-based \sorec}: To the best of our knowledge, we are the first to systematically dedicate ourselves to reviewing GNN-based \sorec methods. Most of the existing surveys focus either on traditional methods~\cite{PapadimitriouSM12,TangHL13,YangGLS14,XuWZC15,DouYD16,ChenHCWZK18,ShokeenR18} (\eg matrix factorization), feature information~\cite{shokeen2020studyFeatures} (\eg context), or a specific application~\cite{GasparettiSM21} (\eg community detection).
    The other related surveys~\cite{Deng22,wangHW0SOC0Y21,wu2020gnnsurvey,gao2021gnnsurvey} focus on graph-based recommender systems, but they partially cover \sorec.
    \item \textbf{Comprehensive Survey}: We systematically identify the relevant papers on GNN-based \sorec by following the guidelines of the preferred reporting items for systematic reviews and meta-analyses (PRISMA framework)~\cite{moher2009preferred}. Then, we comprehensively review them in terms of their inputs and architectures. Figure~\ref{fig:timeline} provides a brief timeline of GNN-based \sorec methods. In addition, Figure~\ref{fig:venue} shows the number of relevant papers published in relevant journals (\eg IEEE TKDE and ACM TOIS) and conferences (\eg WWW, ACM SIGIR, and ACM CIKM).
    \item \textbf{Novel Taxonomy of Inputs and Architectures}: We provide a novel taxonomy of inputs and architectures in GNN-based \sorec methods, enabling researchers to capture the research trends in this field easily. An input taxonomy includes 5 groups of input type notations and 7 groups of input representation notations. On the other hand, an architecture taxonomy includes 8 groups of GNN encoder notations, 2 groups of decoder notations, and 12 groups (4 for primary losses and 8 for auxiliary losses) of loss function notations.
    \item \textbf{Benchmark Datasets}: We review 17 benchmark datasets used to evaluate the performance of GNN-based \sorec methods. We group the datasets into 8 domains (\ie product, location, movie, image, music, bookmark, microblog, and miscellaneous). Also, we present some statistics for each dataset and a list of papers using the dataset.
    \item \textbf{Future Directions}: We discuss the limitations of existing GNN-based \sorec methods and provide several future research directions. 
\end{itemize}

The rest of this survey paper is organized as follows. 
In Section~\ref{sec:survey}, we introduce the survey methodology based on PRISMA~\cite{moher2009preferred} that collects the papers on GNN-based \sorec thoroughly.
In Section~\ref{sec:problem}, we define the social recommendation problem.
In Sections~\ref{sec:inputs} and~\ref{sec:arch}, we review 84 GNN-based \sorec methods in terms of their inputs and architectures, respectively. We summarize 
17 benchmark datasets and 8 evaluation metrics, widely-used in GNN-based \sorec methods, in Section~\ref{sec:setup}.
Section~\ref{sec:directions} discusses future research directions. 
Finally, we conclude the paper in Section~\ref{sec:conclusions}.

\section{Survey Methodology}\label{sec:survey}

Following the guidelines set by the PRISMA~\cite{moher2009preferred}, the Scopus index was queried to filter for relevant literature. In particular, the following query was run on October 14, 2022, resulting in $2,151$ papers.
\begin{center}
{\footnotesize
\texttt{
    \textcolor{red}{TITLE-ABS-KEY} (social \textcolor{blue}{AND} (recommendation \textcolor{blue}{OR} recommender) \textcolor{blue}{AND} graph) \textcolor{blue}{AND} (\textcolor{red}{PUBYEAR} > 2009) \textcolor{blue}{AND} (\textcolor{red}{LIMIT-TO} (\textcolor{red}{LANGUAGE}, ``English''))}}
\end{center}

To obtain the final list of relevant papers for the current survey, an iterative strategy of manual reviewing and filtering was carried out, following PRISMA guidelines. Four expert annotators were used to select the relevant papers. Before reviewing the papers, a comprehensive and exhaustive discussion was held among the annotators to discuss and agree upon the definitions of the main concepts that a paper is to be examined for before including it in the survey. These included concepts of Graph Neural Networks and Social Recommendation. 

Based on these guidelines, each annotator labeled one batch of $200$ papers together. Each paper in this batch was assigned one of the three categories by each annotator: ``Yes'', ``No'', and ``Maybe''. ``Yes'' represents full confidence of relevance, ``Maybe'' represents some confidence of relevance, and ``No'' represents full confidence of irrelevance of the paper for the current survey. A high inter-annotator agreement of $0.845$ among the annotators was reported on this set. 

The remaining papers were then divided equally among the annotators without any overlap. The annotator assigned each paper a label of ``Yes'', ``No'', or ``Maybe''. Papers marked ``Maybe'' were reviewed again by the other annotators to reach a consensus. Finally, papers marked ``Yes'' were collected together and these served as the focus of our survey. Through this comprehensive process of filtering, we finally found $84$ papers that study GNN-based \sorec for our survey paper.

\begin{table}[!t]
\caption{Notations used in this paper.
}
\label{tab:notations}
\resizebox{0.67\textwidth}{!}{
\renewcommand{\arraystretch}{1.5}
\begin{tabular}{c|l}
\toprule
\textbf{Notation} & \multicolumn{1}{c}{\textbf{Description}} \\
\midrule
$\mathcal{U}$, $\mathcal{I}$ & Sets of users $p_i$ and items $q_j$ \\
$\mathbf{R}$, $\mathbf{S}$ & Matrices representing U-I rating and U-U social \\
$\mathcal{N}_{p_i}$ & Set of items rated by $p_i$ \\ \hline
$\mathbf{u}_i^I$ & Embedding of $p_i$ obtained via the user interaction encoder\\ 
$\mathbf{u}_i^S$ & Embedding of $p_i$ obtained via the user social encoder\\
$\mathbf{u}_i$ & Embedding of $p_i$ obtained by fusing $\mathbf{u}_i^I$ and $\mathbf{u}_i^S$ \\
$\mathbf{v}_j$ & Embedding of $q_j$ via the item encoder \\ \hline
$r_{ij}$ & Real rating score of $p_i$ on $q_j$ \\
$\hat{r}_{ij}$ & Predicted preference of $p_i$ on $q_j$ via the decoder \\
\bottomrule
\end{tabular}
}
\end{table}

\section{Notations and Problem Definition}\label{sec:problem}

The social recommendation problem is formulated as follows.
Let $\mathcal{U}=\{p_1,p_2,\cdots,p_m\}$ and $\mathcal{I}=\{q_1,q_2,\cdots,q_n\}$ be sets of $m$ users and $n$ items, respectively. 
Also, $\mathbf{R}\in\mathbb{R}^{m\times n}$ represents a rating matrix that stores user-item ratings (that we call U-I rating).
$\mathbf{S}\in\mathbb{R}^{m\times m}$ represents a social matrix that stores user-user social relations (that we call U-U social).
In addition, $\mathcal{N}_{p_i}$ indicates a set of items rated by a user $p_i$.
In this paper, we use bold uppercase letters and bold lowercase letters to denote matrices and vectors, respectively.
Also, we use calligraphic letters to denote sets and graphs.
Table~\ref{tab:notations} summarizes a list of notations used in this paper.

The goal of GNN-based \sorec methods is to solve the rating prediction and/or top-$N$ recommendation tasks. 
Given $\mathbf{R}$ and $\mathbf{S}$, both tasks are formally defined as follows:

\begin{prob}[\textbf{Rating Prediction}]
    \label{task:rp}
    The goal is to predict the rating values for unrated items (\ie $\mathcal{I}$\textbackslash$\mathcal{N}_{p_i}$) in $\mathbf{R}$ as close as possible to the ground truth.
\end{prob}

\begin{prob}[\textbf{Top-\boldsymbol{$N$} Recommendation}]
    \label{prob:topn}
    The goal is to recommend the top-$N$ items that are most likely to be preferred by each user $p_i$ among $p_i$'s unrated items (\ie $\mathcal{I}$\textbackslash$\mathcal{N}_{p_i}$). 
\end{prob}

\section{Taxonomy of Inputs}\label{sec:inputs}

\begin{figure*}[t]
  \centering
  \includegraphics[width=\linewidth]{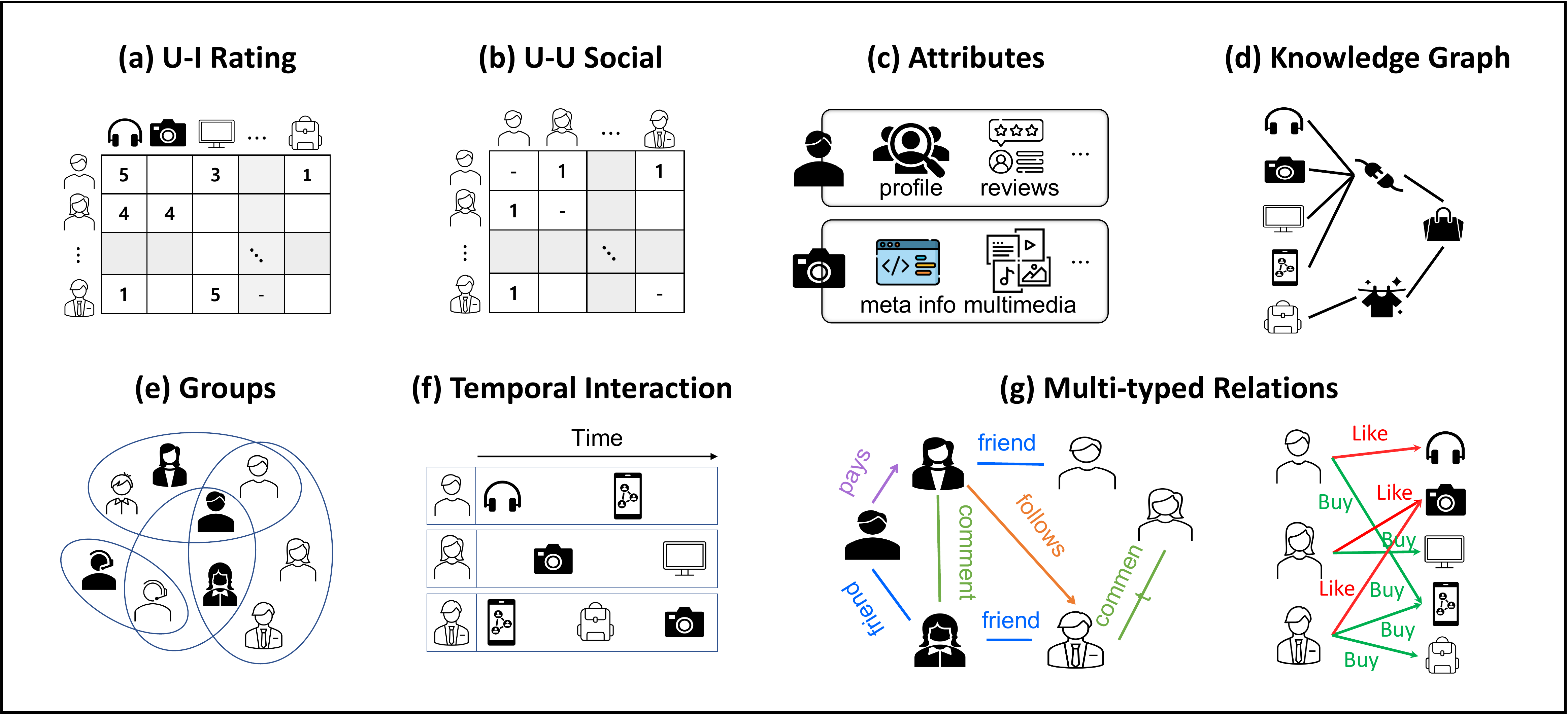}
  \caption{Overview of input types used by GNN-based \sorec methods. }
  \label{fig:input_type}
\end{figure*}

\begin{figure*}[t]
  \centering
  \includegraphics[width=\linewidth]{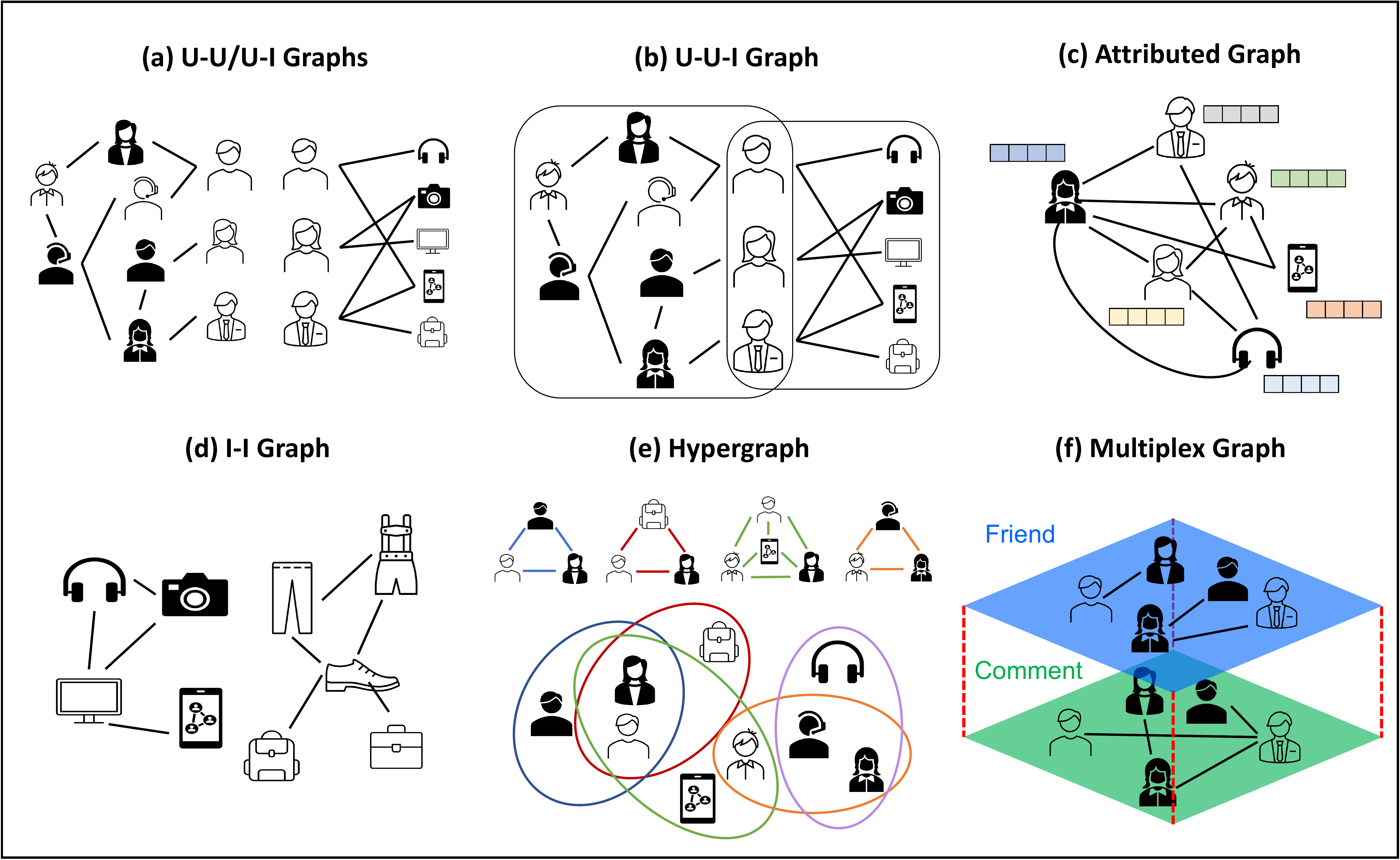}
  \caption{Overview of input representations used by GNN-based \sorec methods.}
  \label{fig:inputs_representation}
\end{figure*}

In this section, we present a taxonomy of inputs for GNN-based \sorec. 
Figures~\ref{fig:input_type} and~\ref{fig:inputs_representation} depict the input types and their representations, respectively.
In the subsequent subsections, we describe each of these in detail.

\subsection{Input Types: Types of Inputs to the Models}
In this subsection, we group the input types used by GNN-based \sorec into 5 categories: user-item ratings, user-user social relations, attributes, knowledge graph (KG), and groups.
Table~\ref{tab:inputs} categorizes all papers based on the input data types they use.

\begin{table*}[t]
    \centering
    \caption{Taxonomy of input types.} 
    \label{tab:inputs}
    \renewcommand{\arraystretch}{2.0}
    \renewcommand{\aboverulesep}{0.0pt}
    \renewcommand{\belowrulesep}{0.0pt}
    \renewcommand{\belowbottomsep}{1.722pt}
    \renewcommand{\abovetopsep}{2.798pt}
    \resizebox{\textwidth}{!}{
    \begin{tabular}{c|c|c|c|l}
    \toprule
    \multirow{2}{*}{\textbf{U-I Rating}} & \multirow{2}{*}{\textbf{U-U Social}} & \multicolumn{2}{c|}{\textbf{Additional Features}} & \multicolumn{1}{c}{\multirow{2}{*}{\textbf{Models}}} \\ \cmidrule{3-4}
     &  & \textbf{User} & \textbf{Item} &  \\
    \midrule
    \multirow{15}{*}{Static} &	\multirow{13}{*}{Homogeneous} &	\multirow{9}{*}{-} & \multirow{7}{*}{-}	 &	GraphRec~\cite{fan2019graphrec}, DANSER~\cite{wu2019danser}, DICER~\cite{fu2021dicer},  ASR~\cite{jiang2021asr}, GNNTSR~\cite{mandal2021gnntsr}, GAT-NSR~\cite{mu2019gatnsr},   \\ 
    & & & & SoRecGAT~\cite{vijaikumar2019sorecgat}, SAGLG~\cite{liu2021sagclg}, PA-GAN~\cite{hou2021pagan}, MGNN~\cite{xiao2020mgnn}, MutualRec~\cite{xiao2021mutualrec},  \\ 
    & & & &   GHSCF~\cite{bi2021ghscf}, HIDM~\cite{li2020hidm}, SocialLGN~\cite{liao2022sociallgn}, SOAP-VAE~\cite{walker2021soapvae}, GraphRec+~\cite{fan2022graphrecp}, DSR~\cite{sha2021dsr}, \\ 
    & & & & SHGCN~\cite{zhu2021shgcn}, GTN~\cite{hoang2021gtn}, PDARec~\cite{zheng2021pdarec}, MHCN~\cite{yu2021mhcn}, SEPT~\cite{yu2021sept}, DcRec~\cite{wu2022dcrec},\\
    & & & &  GSFR~\cite{xiao2022gsfr}, APTE~\cite{zhen2022apte}, EAGCN~\cite{wu2022eagcn}, HOSR~\cite{liu2022hosr}, SDCRec~\cite{du2022sdcrec}, SoHRML~\cite{liu2022sohrml}, DESIGN~\cite{tao2022design} \\
    & & & &   HyperSoRec~\cite{wang2021hypersorec}, CGL~\cite{zhang2022cgl}, DISGCN~\cite{li2022disgcn}, ME-LGN~\cite{miao2022melgn}, GDSRec~\cite{chen2022gdsrec}, SGA~\cite{liufu2021sga}\\
    & & & &   
     SIGA~\cite{liu2022siga},  ESRF~\cite{yu2022esrf}, Motif-Res~\cite{sun2022motifres}, FeSoG~\cite{liu2022fesog}, GDMSR~\cite{gdmsr23}, MADM~\cite{madm24}, DSL~\cite{dsl23}\\ \cmidrule{4-5}
    
    &  &   &	\multirow{1}{*}{KG} &	KConvGraphRec~\cite{tien2020kconvgraphrec}, HeteroGraphRec~\cite{salamat2021heterographrec}, Social-RippleNet~\cite{jiang2022socialripplenet}, SCGRec~\cite{yang2022scgrec} \\ \cmidrule{4-5}
     
     &	 &  &	Attributes  &	GNN-SOR~\cite{guo2020gnnsor}, MPSR~\cite{liu2022mpsr}, FBNE~\cite{chen2022fbne} \\ \cmidrule{3-5}
     
    

     &  & \multirow{2}{*}{Attributes} & \multirow{2}{*}{Attributes} &	Diffnet~\cite{wu2019diffnet}, Diffnet++~\cite{wu2020diffnet++}, DiffnetLG~\cite{song2021diffnetlg}, IDiffNet~\cite{li2022idiffnet}, MEGCN~\cite{jin2020megcn}, SAN\cite{jiang2021san}, \\ 
    & & & &  SRAN~\cite{xie2022sran}, MrAPR~\cite{song2022mrapr}, SENGR~\cite{shi2022sengr}, TAG~\cite{qiao2022tag}, ATGCN~\cite{seng2021atgcn}, HSGNN~\cite{wei2022hsgnn} \\ \cmidrule{3-5}
    
    
     &	&  \multirow{2}{*}{Groups}	 &	- &	IGRec~\cite{chen2022igrec}, GLOW~\cite{leng2022glow} \\ 
    
     &	&  &	Attributes &	GMAN~\cite{liao2022gman} \\ \cmidrule{2-5}

    
      & \multirow{2}{*}{Multiple} & - & Attributes & DH-HGCN~\cite{han2022dhhgcn} \\ 
      &  & Attributes & Attributes & BFHAN~\cite{zhao2021bfhan} \\ \midrule
     
    \multirow{5}{*}{Temporal} &	\multirow{5}{*}{Homogeneous} & \multirow{4}{*}{-} &	\multirow{3}{*}{-} & EGFRec~\cite{gu2021egfrec}, FuseRec~\cite{narang2021fuserec}, DGARec-R~\cite{sun2020dgarecr}, MGSR~\cite{niu2021mgsr}, MOHCN~\cite{wang2022mohcn}, \\
    & & & & GNNRec~\cite{liu2022gnnrec}, DCAN~\cite{wang2022dcan}, SGHAN~\cite{wei2022sghan}, SSRGNN~\cite{chen2022ssrgnn}, GNN-DSR~\cite{lin2022gnndsr}, \\
    & & & & DGRec~\cite{song2019dgrec}, DREAM~\cite{song2020dream}, SeRec~\cite{chen2021serec}, TGRec~\cite{bai2020tgnn} \\ \cmidrule{4-5}
    
    
     &	 &  &	KG &	SSDRec~\cite{yan2022ssdrec} \\ \midrule
     
    Multiple & Homogeneous & - & - & SR-HGNN~\cite{xu2020srhgnn} \\ 
    
    \bottomrule
    \end{tabular}
    }
\end{table*}

\subsubsection{\textbf{User-Item Rating}}
Users interact with different items as they rate them, thus forming the rating matrix $\mathbf{R}\in \mathbb{R}^{m \times n}$. Therefore, each user has a list of items that he/she has interacted with along with the corresponding rating. 

The timestamp of the user-item interaction may also be available and can be exploited to recommend items to users at specific points in time. 
Each rating can thus also be associated with a timestamp for that rating. 
Some models exploit the temporal information to make more-effective recommendations in continuous time~\cite{bai2020tgnn,sun2020dgarecr,narang2021fuserec} or during a user session~\cite{yan2022ssdrec,gu2021egfrec,song2019dgrec,song2020dream}. 

Furthermore, one may also have multi-typed user-item interactions. For example, a user may interact with an item positively (positive rating) or negatively (negative rating). Some models have distinguished among these different interaction types to predict each type more effectively~\cite{xu2020srhgnn}.

\subsubsection{\textbf{User-User Social}}

The second essential input to \sorec is the social adjacency matrix $\mathbf{S} \in \mathbb{R}^{m \times m}$, storing user-user social relations. 

People may be connected to each other via different kinds of social relations. For example, two users may be related if they are friends or if they may 
co-comment on an item or if one follows the other, etc. DH-HGCN~\cite{han2022dhhgcn} and BFHAN~\cite{zhao2021bfhan} consider multifaceted, heterogeneous user-user relations in the social network.

\subsubsection{\textbf{Additional Features}}
\hfill \break
\vspace{1mm}
\indent\textbf{Attributes.}
Both user and items may have additional attributes that can be encoded by the models to make better social recommendations. User attributes are often features of user profiles on social media, \eg age, sex, etc., while item attributes are often information about the items such as its price and category. Some models just incorporate user attributes~\cite{xiao2021mutualrec,seng2021atgcn}, some only item attributes~\cite{guo2020gnnsor,chen2021serec}, and others incorporate both~\cite{wu2019diffnet,wu2020diffnet++,song2021diffnetlg,jin2020megcn,jiang2021san}.

\vspace{1mm}
\textbf{Knowledge Graph (KG).}
Items are structured on a product site in the form of a knowledge graph where items are related with each other if they have some mutual dependency. Models incorporate such dependencies between items as represented by this knowledge graph~\cite{tien2020kconvgraphrec,salamat2021heterographrec}. 

\vspace{1mm}
\textbf{Groups.}
Users are often grouped together denoting a group structure among them. For instance, multiple users can form an online social group based on similar interests or hobbies. Models incorporate the group membership in addition to the social relations to model the social network more effectively~\cite{liao2022gman,chen2022igrec,leng2022glow}. User groups can also be formed based on the businesses that they are part of or are clients of, as in ~\cite{chen2022fbne}. 

\begin{table*}[t]
    \centering
    \caption{Taxonomy of input representations.}\label{tab:inputreps}
    \renewcommand{\arraystretch}{1.8}
    \renewcommand{\aboverulesep}{0.0pt}
    \renewcommand{\belowrulesep}{0.0pt}
    \renewcommand{\belowbottomsep}{0.4ex}
    \renewcommand{\abovetopsep}{0.65ex}
    \resizebox{\textwidth}{!}{
    \begin{tabular}{c|l}
    \toprule
        \textbf{Graph Representations} & \multicolumn{1}{c}{\textbf{Models}} \\
        \midrule
        \multirow{8}{*}{U-U/U-I} & GraphRec~\cite{fan2019graphrec}, DANSER~\cite{wu2019danser}, DICER~\cite{fu2021dicer},  ASR~\cite{jiang2021asr}, GNNTSR~\cite{mandal2021gnntsr}, GAT-NSR~\cite{mu2019gatnsr}, \\
        & SoRecGAT~\cite{vijaikumar2019sorecgat}, SAGLG~\cite{liu2021sagclg}, PA-GAN~\cite{hou2021pagan}, MGNN~\cite{xiao2020mgnn}, MutualRec~\cite{xiao2021mutualrec}, GHSCF~\cite{bi2021ghscf}, \\
        & HIDM~\cite{li2020hidm}, SocialLGN~\cite{liao2022sociallgn}, SOAP-VAE~\cite{walker2021soapvae}, GraphRec+~\cite{fan2022graphrecp}, DSR~\cite{sha2021dsr}, GTN~\cite{hoang2021gtn}, \\
        & PDARec~\cite{zheng2021pdarec}, SEPT~\cite{yu2021sept}, DcRec~\cite{wu2022dcrec}, GSFR~\cite{xiao2022gsfr}, APTE~\cite{zhen2022apte}, EAGCN~\cite{wu2022eagcn}, \\
        & HOSR~\cite{liu2022hosr}, SDCRec~\cite{du2022sdcrec}, SoHRML~\cite{liu2022sohrml}, DESIGN~\cite{tao2022design}, HyperSoRec~\cite{wang2021hypersorec}, CGL~\cite{zhang2022cgl}, \\
        & DISGCN~\cite{li2022disgcn}, GDSRec~\cite{chen2022gdsrec}, SGA~\cite{liufu2021sga}, SIGA~\cite{liu2022siga},  ESRF~\cite{yu2022esrf},  SR-HGNN~\cite{xu2020srhgnn}, \\
        & EGFRec~\cite{gu2021egfrec}, FuseRec~\cite{narang2021fuserec}, DGARec-R~\cite{sun2020dgarecr}, MGSR~\cite{niu2021mgsr}, GNNRec~\cite{liu2022gnnrec}, DCAN~\cite{wang2022dcan}, SGHAN~\cite{wei2022sghan} \\
        &  GNN-DSR~\cite{lin2022gnndsr}, DGRec~\cite{song2019dgrec}, DREAM~\cite{song2020dream}, SeRec~\cite{chen2021serec}, TGRec~\cite{bai2020tgnn}, MADM~\cite{madm24}, DSL~\cite{dsl23}\\
        \midrule
        U-U-I & SHGCN~\cite{zhu2021shgcn}, SSRGNN~\cite{chen2022ssrgnn}, ME-LGN~\cite{miao2022melgn}, SENGR~\cite{shi2022sengr}, IGRec~\cite{chen2022igrec}, GLOW~\cite{leng2022glow}, GDMSR~\cite{gdmsr23} \\
        \midrule
        \multirow{3}{*}{Attributed} & DiffNet~\cite{wu2019diffnet}, DiffNet++~\cite{wu2020diffnet++}, DiffNet-LG~\cite{song2021diffnetlg}, IDiffNet~\cite{li2022idiffnet}, MEGCN~\cite{jin2020megcn}, SAN~\cite{jiang2021san}, \\
        & SRAN~\cite{xie2022sran}, MrAPR~\cite{song2022mrapr}, SENGR~\cite{shi2022sengr}, TAG~\cite{qiao2022tag}, ATGCN~\cite{seng2021atgcn}, HSGNN~\cite{wei2022hsgnn}, \\
        & GMAN~\cite{liao2022gman}, GNN-SOR~\cite{guo2020gnnsor}, MPSR~\cite{liu2022mpsr}, FBNE~\cite{chen2022fbne}, DH-HGCN~\cite{han2022dhhgcn}, BFHAN~\cite{zhao2021bfhan} \\
        \midrule
        Multiplex & DH-HGCN~\cite{han2022dhhgcn}, BFHAN~\cite{zhao2021bfhan}\\
        \midrule
        \multirow{2}{*}{U-U/U-I/I-I} & KConvGraph~\cite{tien2020kconvgraphrec}, HeteroGraphRec~\cite{salamat2021heterographrec}, Social-RippleNet~\cite{jiang2022socialripplenet}, SCGRec~\cite{yang2022scgrec}, \\
        & SSDRec~\cite{yan2022ssdrec}, DGNN~\cite{dgnn23} \\
        \midrule
        Hypergraph & DH-HGCN~\cite{han2022dhhgcn}, MHCN~\cite{yu2021mhcn}, SHGCN~\cite{zhu2021shgcn}, Motif-Res~\cite{sun2022motifres}, MOHCN~\cite{wang2022mohcn}\\
        \midrule
        Decentralized & FeSoG~\cite{liu2022fesog} \\
    \bottomrule
    \end{tabular}
    }
\end{table*}

\subsection{Input Representations: Representation of Inputs within the Models}
In order to effectively use the available inputs with GNN-based models, \sorec methods represent them as different graphs. 
In particular, the input representations employed by GNN-based \sorec can be grouped into 7 categories: U-U/U-I graphs, U-U-I graph, attributed graph, multiplex graph, U-U/U-I/I-I graphs, hypergraph, and decentralized. 
Table~\ref{tab:inputreps} categorizes papers based on the input representation they develop using the input data.

\subsubsection{\textbf{U-U/U-I Graphs}}
The simplest representation of the input for social recommendation is to use separate graphs for a user-user social network and a user-item interaction network. The user-item interaction network is represented as a bipartite graph and the user-user social network is represented as a general undirected/directed graph. Information from the two graphs is encoded separately at the common user node and later aggregated. Most works follow this representation to encode users and items~\cite{fan2019graphrec,wu2019diffnet,fu2021dicer,wu2019danser,fan2022graphrecp,xiao2021mutualrec,xiao2020mgnn,li2020hidm,wu2020diffnet++,song2021diffnetlg,tao2022design,wu2022eagcn}.

\subsubsection{\textbf{U-U-I Graph}}
Both kinds of user-user relations and user-item interactions can be modeled together by a single graph as well. Here, user-user edges and user-item edges in the graph need to be distinguished by the type of the end node. Many works thus merge the social relation edges and interaction edges together in a single graph to obtain node embeddings for both users and items~\cite{zhu2021shgcn,chen2022ssrgnn,miao2022melgn,shi2022sengr}. 

\subsubsection{\textbf{Attributed Graph}}
Both user and item nodes may further contain features describing the corresponding entity. For example, users may have profile features while items may have their description features. These features are first encoded numerically and then represented explicitly as node attributes in the U-U/U-I graph or U-U-I graph to make effective recommendations~\cite{wu2019diffnet,wu2020diffnet++,song2021diffnetlg,xie2022sran,jiang2021san,song2022mrapr,seng2021atgcn}. These attributes are either fused with the learned embeddings or are used as initialization for the GNN layers. 

\subsubsection{\textbf{Multiplex Graph}}
Users may be related to each other via multiple relationships while they may also interact with items in multiple ways. Such relationships are often represented using a multiplex network, \ie using multiple layers of the U-U/U-I graph, where each layer represents a particular relation type~\cite{xu2020srhgnn,han2022dhhgcn,zhao2021bfhan}. 

\subsubsection{\textbf{U-U/U-I/I-I Graphs}}
When information on item-item relations is available, an item-item knowledge graph is considered in addition to the U-U and U-I graphs. Item embeddings are now obtained separately one from the U-I interaction graph and the other from the item-item knowledge graph and then aggregated later to obtain the final item embedding ~\cite{tien2020kconvgraphrec,salamat2021heterographrec}.

\subsubsection{\textbf{Hypergraph}}
One may want to incorporate higher-order relations among users and items to explicitly establish organizational properties in the input such as (1) constructing a user-only hyperedge if a group of users are connected together in closed motifs, (2) constructing a user-item joint hyperedge if a group of users interacts with the same item, and (3) constructing an item-item hyperedge if one user interacts with a group of items. Models have been developed to include just user-item joint hyperedges~\cite{zhu2021shgcn}, both user-user and user-item joint hyperedges~\cite{yu2021mhcn}, and user-user and item-item hyperedges~\cite{han2022dhhgcn}. 

\subsubsection{\textbf{Decentralized}} 
Centralized data storage is becoming infeasible in practice due to rising privacy concerns. Thus, instead of storing the complete U-U/U-I graphs together, a decentralized storage of the graphs is often required. Here, the edges for social relations and interactions of each user are stored locally at each user's local server such that only non-sensitive data is shared with the centralized server~\cite{liu2022fesog}

\section{Taxonomy of Architectures}\label{sec:arch}

\begin{figure*}[t]
\centering
  \includegraphics[width=\linewidth]{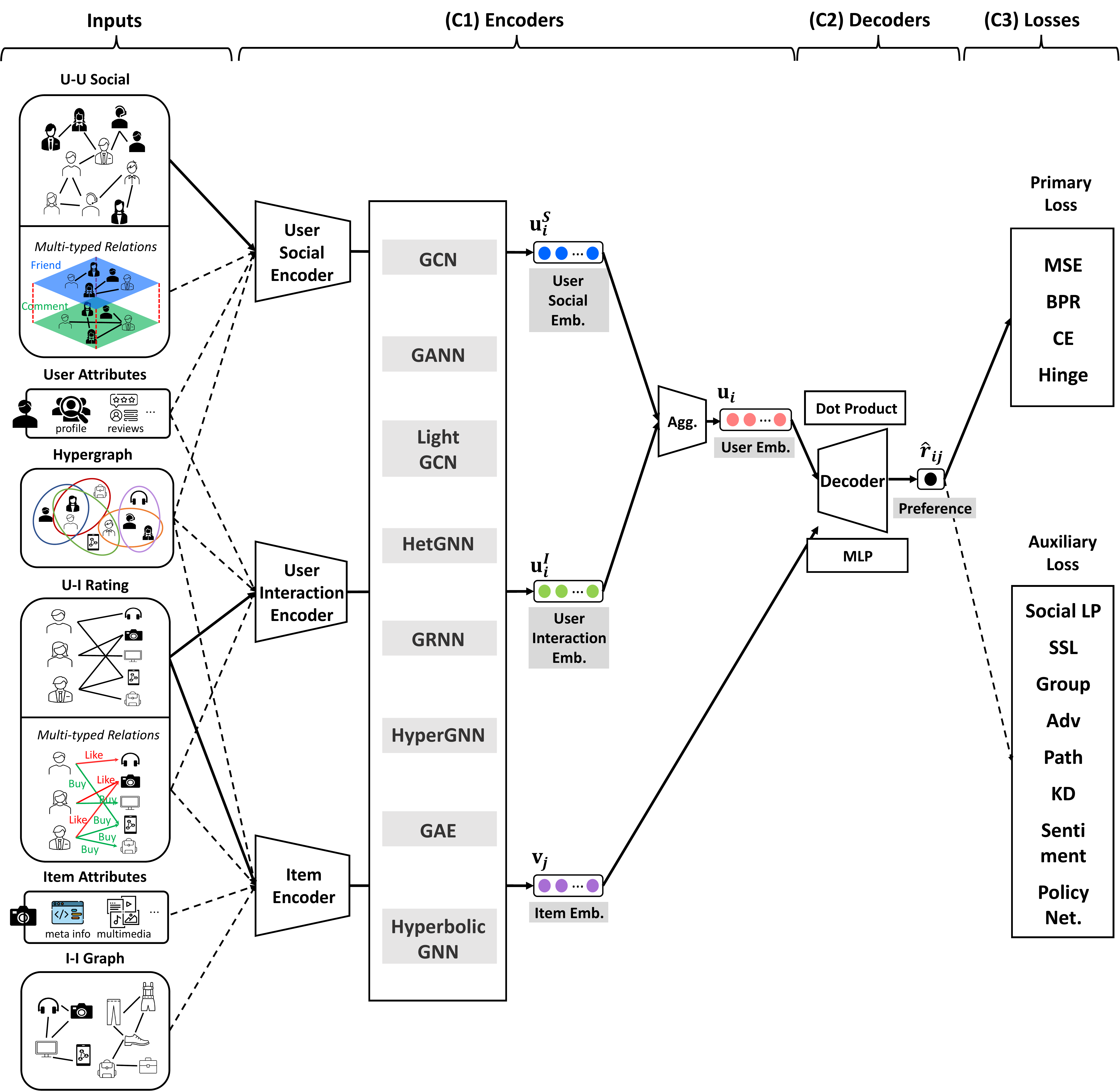}
\caption{Overview of architectures for GNN-based \sorec methods. 
The solid lines represent the common procedure of the GNN-based SocialRS methods, while the dashed lines represent the flows for the auxiliary inputs or the losses.
}
\label{fig:overview}
\end{figure*}

In this section, we present the taxonomy of architectures for GNN-based \sorec.
Model architectures consist of three key components as shown in Figure~\ref{fig:overview}: 
(C1) encoders; (C2) decoders; (C3) loss functions.
Using the U-U and U-I graphs in (C1), the encoders create low-dimensional vectors (\mbox{\ie} embeddings) for users and items by employing different GNN encoders.
Here, some works exploit additional information of users and/or items (\eg their attributes and groups; refer to Section~\ref{sec:inputs}) to construct more-accurate user and item embeddings.
In Figure~\mbox{\ref{fig:overview}}, the dashed lines show which encoders use each additional piece of information.
In (C2), the decoders predict each user's preference on each item via different operations on the user and item embeddings obtained from (C1). 
Finally, in (C3), different loss functions are optimized to learn the embeddings in an end-to-end manner.
We discuss the advantages and disadvantages of each encoder in Table~\mbox{\ref{tab:comp_encoders}} while the loss functions are discussed in Table~\mbox{\ref{tab:comp_loss}}.
In the subsequent subsections, we describe each component of GNN-based \sorec in detail.

\begin{table}[t]
    \centering
    \caption{High-level comparison of different encoders.}
    \renewcommand{\arraystretch}{1.2}
    \renewcommand{\aboverulesep}{0.0pt}
    \renewcommand{\belowrulesep}{0.0pt}
    \renewcommand{\belowbottomsep}{1.722pt}
    \renewcommand{\abovetopsep}{2.798pt}
    \newcolumntype{H}{>{\setbox0=\hbox\bgroup}c<{\egroup}@{}}
    \resizebox{\textwidth}{!}{
    \begin{tabular}{ccccc}
    \toprule
    \textbf{Encoder} & \textbf{Complexity} & \textbf{Representativeness} & \textbf{Additional Features} & \textbf{Known Issues} \\
    \midrule
    GCN  &  Low & Low & None & Oversmoothing \\
    LightGCN & Very Low & Lower & None & Linear representation \\
    GANN  & High & High & None & Oversmoothing, more parameters \\
    HetGNN & High & Higher & Heterogeneous interactions & Extra annotation, Hard to \\
    & & & & generalize over interactions \\
    GRNN & Very High & Higher & Temporal interactions & Vanishing and exploding gradients\\
    HyperGNN & Higher & Higher & High-order relations & Extra motif annotation\\
    GAE & Very High & High & Generative & Harder to optimize \\
    Hyperbolic GNN & High & Very High & Hierarchical nature & Less empirical evidence  \\
    \bottomrule
    \end{tabular}}
    \label{tab:comp_encoders}
\end{table}
\begin{table}[t]
    \centering
    \caption{High-level comparison of different loss functions.}
    \renewcommand{\arraystretch}{1.2}
    \renewcommand{\aboverulesep}{0.0pt}
    \renewcommand{\belowrulesep}{0.0pt}
    \renewcommand{\belowbottomsep}{1.722pt}
    \renewcommand{\abovetopsep}{2.798pt}
    \newcolumntype{H}{>{\setbox0=\hbox\bgroup}c<{\egroup}@{}}
    \resizebox{\textwidth}{!}{
    \begin{tabular}{ccccc}
    \toprule
    \textbf{Loss function} & \textbf{Complexity} & \textbf{Benefits} & \textbf{Known Issues} \\
    \midrule
    MSE & Low & Learns continuous ratings & Sensitive to outliers\\
    BPR & Low & Learns rankings between items & Cannot handle continuous ratings\\
    CE & Low & Suited for classification & Cannot handle continuous ratings \\
    Hinge & Low & Faster convergence & Cannot handle continuous ratings \\
    \midrule
    Social LP & Low & More suitable social embeddings & Multi-objective trade-off \\
    SSL & High & More informed representations & Expensive pre-processing \\
    Group & Low & Group-level representations & Groups form for specific items \\
    & & & (common interests)\\
    Adv & High & More robust representations & Adversarial instability and cost\\
    Path & High & Predicts social influence propagation & Complicated auxiliary task \\
    KD & High & Less overfitting & Training multiple models\\
    Sentiment & High & User sentiment-weighed ratings & Complex sentiment classification task\\
    Policy Net. & High & Importance weights to each component & Expensive weight allocation \\
    \bottomrule
    \end{tabular}}
    \label{tab:comp_loss}
\end{table}


\begin{table}[t]
    \centering
    \caption{Taxonomy of encoder architectures. It should be noted that some methods employ non-GNN encoders (\eg RNN, MLP, or just an embedding vector (Emb)) or no encoders (\ie $-$) to obtain embeddings.}
    \label{tab:encoder}
    \resizebox{\textwidth}{!}{
    \renewcommand{\arraystretch}{2.0}
    \renewcommand{\aboverulesep}{0.0pt}
    \renewcommand{\belowrulesep}{0.0pt}
    \renewcommand{\belowbottomsep}{1.722pt}
    \renewcommand{\abovetopsep}{2.798pt}
    \newcolumntype{H}{>{\setbox0=\hbox\bgroup}c<{\egroup}@{}}
    \begin{tabular}{c|c|c|l}
        \toprule
        \textbf{User Social} & \textbf{User Interest} & \textbf{Item Encoder} & \multicolumn{1}{c}{\textbf{Models}} \\
        \midrule 
        \multirow{14}{*}{GANN} & \multirow{9}{*}{GANN} & \multirow{5}{*}{GANN} &  GraphRec~\cite{fan2019graphrec}, DiffNet++~\cite{wu2020diffnet++}, DiffNetLG~\cite{song2021diffnetlg}, MutualRec~\cite{xiao2021mutualrec}, SR-HGNN~\cite{xu2020srhgnn}, ASR~\cite{jiang2021asr}, \\
        & & &  GNNTSR~\cite{mandal2021gnntsr}, GAT-NSR~\cite{mu2019gatnsr}, SoRecGAT~\cite{vijaikumar2019sorecgat}, PA-GAN~\cite{hou2021pagan}, SOAP-VAE~\cite{walker2021soapvae}, GTN~\cite{hoang2021gtn}, \\
        & & &  PDARec~\cite{zheng2021pdarec}, FeSoG~\cite{liu2022fesog}, ESRF~\cite{yu2022esrf}, SoHRML~\cite{liu2022sohrml}, FBNE~\cite{chen2022fbne}, DISGCN~\cite{li2022disgcn}, ME-LGN~\cite{miao2022melgn}, \\
        & & &  SGA~\cite{liufu2021sga}, GDSRec~\cite{chen2022gdsrec}, DANSER~\cite{wu2019danser}, GraphRec+\cite{fan2022graphrecp},  SRAN~\cite{xie2022sran}, TAG~\cite{qiao2022tag}, SDCRec~\cite{du2022sdcrec}, \\
         &  &  &  KConvGraph~\cite{tien2020kconvgraphrec}, HeteroGraphRec~\cite{salamat2021heterographrec}, SocialRippleNet~\cite{jiang2022socialripplenet}, SCGRec~\cite{yang2022scgrec}, TGRec~\cite{bai2020tgnn}, \\
         & & & GSFR~\cite{xiao2022gsfr}, IGRec~\cite{chen2022igrec}, DICER~\cite{fu2021dicer} \\ \cmidrule{3-4}
         &  & \multirow{1}{*}{Emb} &  SAN~\cite{jiang2021san}, HIDM~\cite{li2020hidm}, GLOW~\cite{leng2022glow}, GMAN~\cite{liao2022gman}, HSGNN~\cite{wei2022hsgnn}, MrAPR~\cite{song2022mrapr} \\ \cmidrule{3-4}
         &  & GCN & BFHAN~\cite{zhao2021bfhan}, SHGCN~\cite{zhu2021shgcn} \\ \cmidrule{2-4}
         
         & \multirow{2}{*}{GRNN} & GRNN & DGARec-R~\cite{sun2020dgarecr}, SSRGNN~\cite{chen2022ssrgnn}, GNN-DSR~\cite{lin2022gnndsr} \\
         &  & Emb &  DREAM~\cite{song2020dream} \\ \cmidrule{2-4}
         & \multirow{3}{*}{RNN} & GANN & FuseRec~\cite{narang2021fuserec} \\
         &  & RNN  & SGHAN~\cite{wei2022sghan} \\
         &  & Emb &  SSDRec~\cite{yan2022ssdrec} \\ \cmidrule{2-4}
         & - & - & GHSCF~\cite{bi2021ghscf} \\ \midrule

        \multirow{7}{*}{GCN} & \multirow{2}{*}{GCN} & Emb & DiffNet~\cite{wu2019diffnet}, MEGCN~\cite{jin2020megcn}, MPSR~\cite{liu2022mpsr}, HOSR~\cite{liu2022hosr}, \\ \cmidrule{3-4} 
         &  & \multirow{1}{*}{GCN} & ATGCN~\cite{seng2021atgcn}, SENGR~\cite{shi2022sengr}, SAGLG~\cite{liu2021sagclg}, GNN-SOR~\cite{guo2020gnnsor}, GDMSR~\cite{gdmsr23}, MADM~\cite{madm24}, DSL~\cite{dsl23} \\ \cmidrule{2-4}
         & \multirow{3}{*}{GRNN} & GRNN &  GNNRec~\cite{liu2022gnnrec}, MGSR~\cite{niu2021mgsr} \\ 
         &  & GCN & EGFRec~\cite{gu2021egfrec} \\
         &  & Emb & DGRec~\cite{song2019dgrec} \\ \cmidrule{2-4}
         & RNN & GANN & MOHCN~\cite{wang2022mohcn} \\
         & MLP & Emb & MGNN~\cite{xiao2020mgnn} \\ \midrule

         \multirow{3}{*}{LightGCN} & \multirow{3}{*}{LightGCN} & \multirow{2}{*}{LightGCN} & SocialLGN~\cite{liao2022sociallgn}, DcRec~\cite{wu2022dcrec}, APTE~\cite{zhen2022apte}, EAGCN~\cite{wu2022eagcn}, CGL~\cite{zhang2022cgl} \\
         &  &  &  SEPT~\cite{yu2021sept}, DSR~\cite{sha2021dsr}, DESIGN~\cite{tao2022design} \\ \cmidrule{3-4}
         & & Emb &  IDiffNet~\cite{li2022idiffnet} \\ \midrule

        \multirow{2}{*}{HetGNN} & \multirow{2}{*}{HetGNN} & HetGNN  & SeRec~\cite{chen2021serec}, DGNN~\cite{dgnn23} \\
         &  & GANN  & DCAN~\cite{wang2022dcan} \\ \midrule

        \multirow{1}{*}{HyperGNN} & \multirow{1}{*}{GANN} & \multirow{1}{*}{HyperGNN} &  MHCN~\cite{yu2021mhcn}, Motif-Res~\cite{sun2022motifres}, DH-HGCN~\cite{han2022dhhgcn} \\ \midrule
        
        
        GAE & GAE & GAE &  SIGA~\cite{liu2022siga} \\ \midrule
        
        Hyperbolic & Hyperbolic & Hyperbolic & HyperSoRec~\cite{wang2021hypersorec} \\
        
        \bottomrule
    \end{tabular}}
\end{table}

\subsection{Encoders}
\begin{figure*}[t]
\centering
  \includegraphics[width=\linewidth]{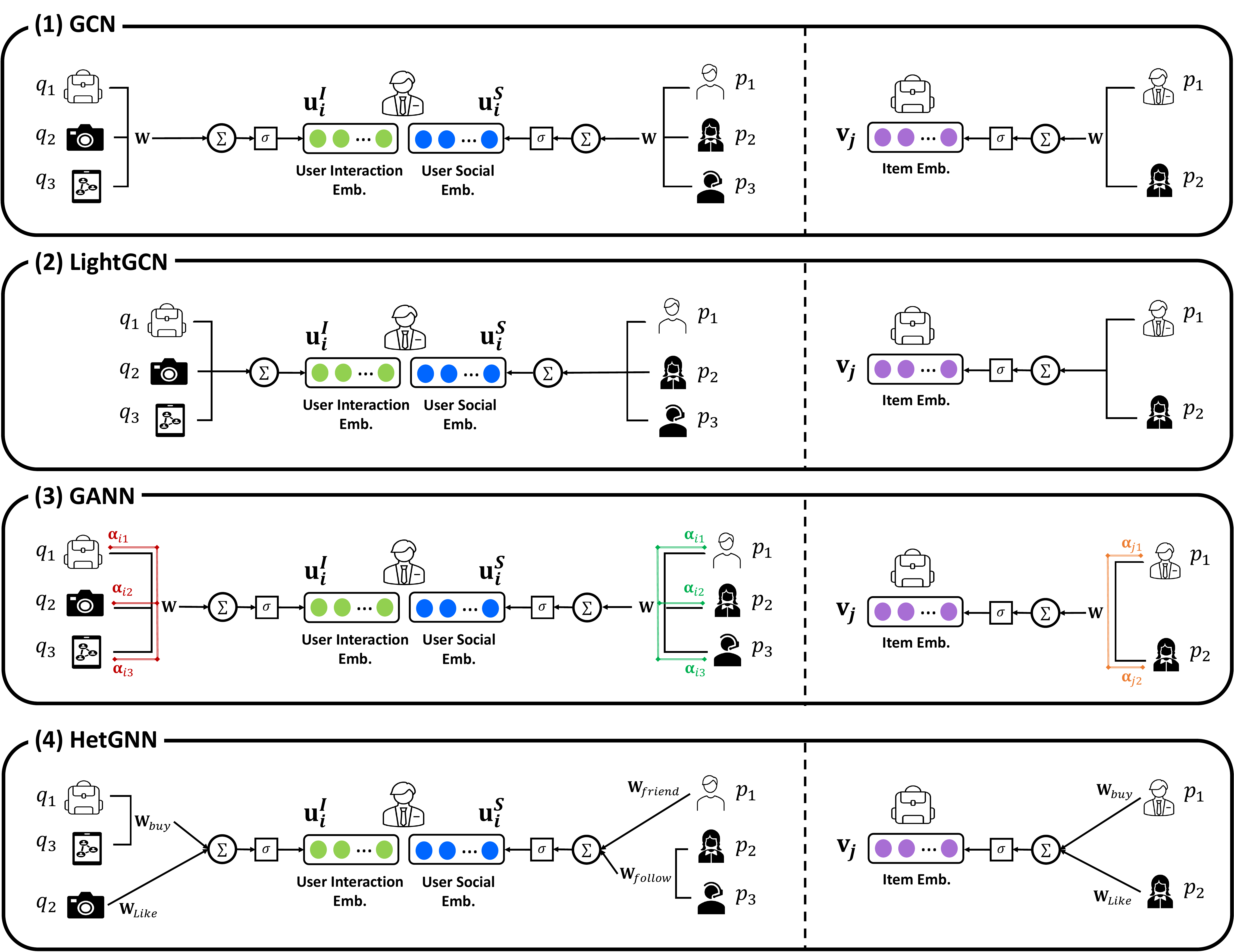}
\caption{Conceptual views of the GCN, LightGCN, GANN, and HetGNN encoders with a single GNN layer. The left side represents user embeddings, while the right side represents item embeddings.}
\label{fig:encoders_basic}
\end{figure*}

\begin{figure*}[t]
\centering
  \includegraphics[width=0.85\linewidth]{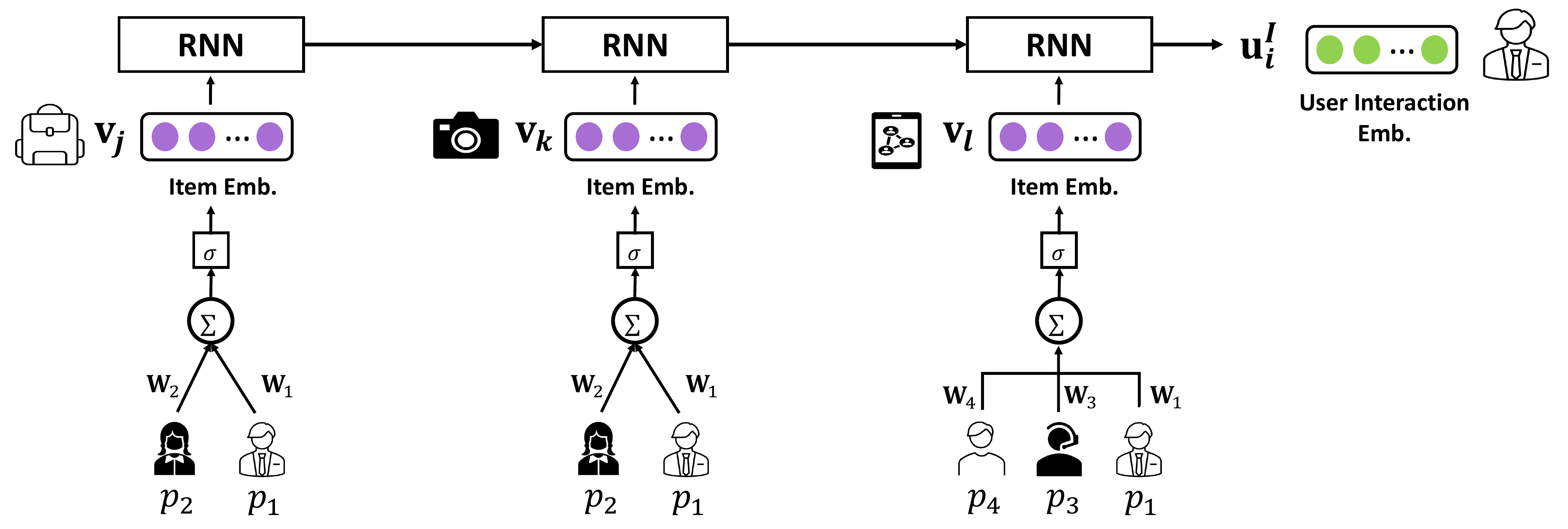}
\caption{Conceptual view of the GRNN encoder with a single GNN layer. Note that existing work did not use GRNN encoders to obtain user social embeddings.}
\label{fig:encoders_grnn}
\end{figure*}

\begin{figure*}[t]
\centering
  \includegraphics[width=0.85\linewidth]{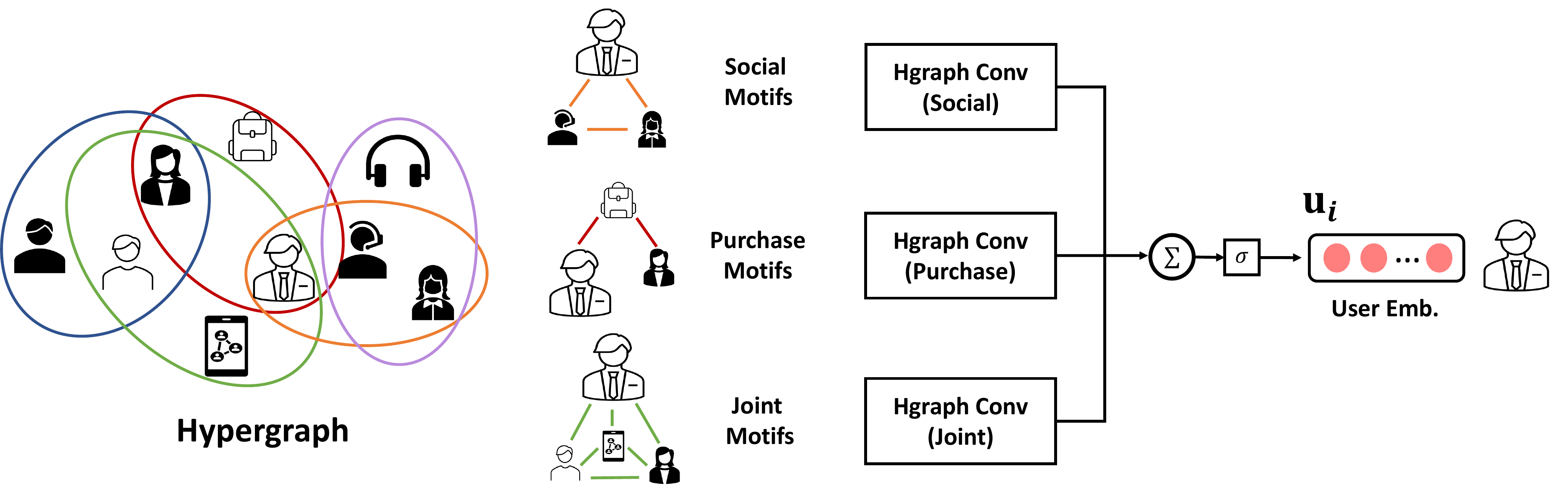}
\caption{Conceptual view of the HyperGNN encoder with a single GNN layer. Note that existing work did not distinguish between user embeddings as interaction and social embeddings, nor did it utilize HyperGNN encoders to generate item embeddings.}
\label{fig:encoders_hypergnn}
\end{figure*}

We group the encoders of GNN-based \sorec into 8 categories: graph convolutional network (GCN), lightweight GCN (LightGCN), graph attention neural networks (GANN), heterogeneous GNN (HetGNN), graph recurrent neural networks (GRNN), hypergraph neural networks (HyperGNN), graph autoencoder (GAE), and hyperbolic GNN.
Table~\ref{tab:encoder} shows the taxonomy of encoders used in existing work in detail.
Figures~\mbox{\ref{fig:encoders_basic}},~\mbox{\ref{fig:encoders_grnn}}, and~\mbox{\ref{fig:encoders_hypergnn}} present the conceptual views for distinct types of GNN encoders.

Generally, in (C1) encoders,
most methods represent each user $p_i$ into two types of low-dimensional vectors (\ie embeddings) by employing a GNN encoder: $p_i$'s interaction embedding $\mathbf{u}_i^I$ based on a U-I graph and $p_i$'s social embedding $\mathbf{u}_i^S$ based on a U-U graph. 
Then, they aggregate them into one embedding $\mathbf{u}_i$ for the corresponding user $p_i$.
In the meantime, they also obtain each item $q_j$'s embedding $\mathbf{v}_j$ via another GNN encoder using a U-I graph.
As mentioned above, some works enhance these embeddings by incorporating additional input representations, such as user/item attributes and hypergraphs.

It should be noted that some works employ only a single GNN encoder to obtain the two embeddings. 
In contrast, others use different GNN encoders for the embeddings of different node types (\ie users or items).
For simplicity, however, we here explain the GNN encoders by generalizing them to any node type in the input graph.

\subsubsection{\textbf{GCN}}
Early works~\cite{wu2019diffnet,jin2020megcn,liu2022mpsr,liu2022hosr,zhu2022sinews,seng2021atgcn,shi2022sengr,liu2021sagclg,guo2020gnnsor,xiao2020mgnn} have focused on representing the user and item embeddings using GCN. 
Given a node $n_i$ (\ie a user or an item) in the input graph (\ie U-I or U-U graphs), a $n_i$'s embedding $\mathbf e_{i}^{k}$ in $k$-th layer is represented based on the embeddings of $n_i$'s neighbors in $(k-1)$-th layer as follows:
\begin{equation}
    \mathbf e_{i}^{(k)} = \sigma(\sum\nolimits_{n_j \in \mathcal{N}_{n_i}} \frac{1}{\sqrt{|\mathcal{N}_{n_j}||\mathcal{N}_{n_i}|}}\mathbf{e}_{j}^{(k-1)} \mathbf{W}^{(k)}),
\end{equation}
where $\sigma$ and $\mathbf{W}^{(k)} \in \mathbb{R}^{d \times d}$ denote a non-linear activation function (\eg ReLU) and a trainable transformation matrix, respectively. Also, $\mathcal{N}_{n_i}$ indicates a set of $n_i$'s neighbors in the input graph.
Here, some works take the self-connection of $n_i$ into consideration by aggregating over the set $\mathcal{N}_{n_i} \cup \{n_i\}$. 
Most methods simply consider the $n_i$'s embedding in the last $K$-th layer, $\mathbf e_{i}^{(K)}$, as its final embedding $\mathbf z_{i}$.
Another variant is to aggregate $n_i$'s embeddings from all layers, \ie $\mathbf z_{i}=\sum^K_{k=1} \mathbf e_{i}^{(k)}$.
For instance, DiffNet~\cite{wu2019diffnet} obtains a user $p_i$'s social embedding $\mathbf u_{i}^S$ (resp. interaction embedding $\mathbf u_{i}^I$) by performing GCN with $k$-layers (resp. $1$-layer) based on the U-U graph (resp. U-I graph).
For each item $q_j$, it simply obtains $q_j$'s embedding $\mathbf v_{j}$ based on its attributes without using a GNN encoder.

Note that different normalization schemes have been proposed in the literature to normalize the weight of each neighbor. The most common strategy is the symmetric normalization as $1/\sqrt{|\mathcal{N}_{n_j}||\mathcal{N}_{n_i}|}$ for its simpler symmetric matrix form. One can also just use $1/|\mathcal{N}_{n_j}|$ but it gives a lower weight to high-degree nodes as compared to the previous form, which may not be desirable. Finally, just using $1/|\mathcal{N}_{n_i}|$ is also not typically desirable as it smooths the neighbor information without considering their degrees. We will thus use symmetric normalization unless otherwise mentioned.

\subsubsection{\textbf{LightGCN}} It is well-known that non-linear activation and feature transformation in GCN encoders make the propagation step very complicated for training and scalability~\cite{he20lightgcn,MaoZXLWH21}.
Motivated by this, some works~\cite{liao2022sociallgn,wu2022dcrec,zhen2022apte,wu2022eagcn,zhang2022cgl,yu2021sept,sha2021dsr,tao2022design,li2022idiffnet} have attempted to replace their GCN encoders with lightweight GCN~\cite{he20lightgcn}, \ie
\begin{equation}
    \mathbf e_{i}^{(k)} = \sum\nolimits_{n_j \in \mathcal{N}_{n_i}} \frac{1}{\sqrt{|\mathcal{N}_{n_j}||\mathcal{N}_{n_i}|}}\mathbf{e}_{j}^{(k-1)}.
\end{equation}
It should be noted that LightGCN~\cite{he20lightgcn} has no non-linear activation function, no feature transformation, and no self-connection.

For instance, DcRec~\cite{wu2022dcrec} obtains each user $p_i$'s social embedding $\mathbf u_{i}^S$ via GCN, whereas obtaining the $p_i$'s interaction embedding $\mathbf u_{i}^I$ and each item $q_j$'s embedding $\mathbf v_{j}$ via the LightGCN encoder.

\subsubsection{\textbf{GANN}} 
The attention mechanism in graphs originated from the graph attention network (GAT)~\cite{velivckovic2017graph} and has already been successful in many applications, including recommender systems. 
Considering different weights from neighbor nodes in the input graph helps focus on important adjacent nodes while filtering out noises during the propagation process~\cite{velivckovic2017graph}.
Therefore, almost all existing works on \sorec have leveraged the attention mechanism in their GNN encoders~\cite{fan2019graphrec,wu2020diffnet++,song2021diffnetlg,xiao2021mutualrec,xu2020srhgnn,jiang2021asr,mandal2021gnntsr,mu2019gatnsr,vijaikumar2019sorecgat,hou2021pagan,walker2021soapvae,hoang2021gtn,zheng2021pdarec,liu2022fesog,yu2022esrf,liu2022sohrml,chen2022fbne,li2022disgcn,miao2022melgn,liufu2021sga,chen2022gdsrec,wu2019danser,fan2022graphrecp,xie2022sran,qiao2022tag,du2022sdcrec,tien2020kconvgraphrec,salamat2021heterographrec,jiang2022socialripplenet,bai2020tgnn,yang2022scgrec,xiao2022gsfr,chen2022igrec,fu2021dicer,chen2022igrec,jiang2021san,li2020hidm,leng2022glow,liao2022gman,wei2022hsgnn,song2022mrapr,zhao2021bfhan,zhu2021shgcn,sun2020dgarecr,chen2022ssrgnn,lin2022gnndsr,song2020dream,narang2021fuserec,wei2022sghan,yan2022ssdrec,bi2021ghscf}.

The common intuitions behind their design of the attention mechanism are:  
(1) each user's preferences for different items may differ, and
(2) each user's influences on her social friends may differ.
Based on such intuitions, many methods represent a node $n_i$'s embedding in $k$-th layer by attentively aggregating the embeddings of $n_i$'s neighbors in $(k-1)$-th layer as follows:
\begin{equation}
    \mathbf e_{i}^{(k)} = \sigma(\sum\nolimits_{n_j \in \mathcal{N}_{n_i}} (\alpha_{ij} \cdot \mathbf{e}_{j}^{(k-1)}) \mathbf{W}^{(k)}),
\end{equation}
where $\alpha_{ij}$ indicates the attention weight of neighbor node $n_j$ w.r.t $n_i$.

Now, we discuss how to compute the attention weights in existing works. 
Most methods, including DANSER~\cite{wu2019danser} and SCGRec~\cite{yang2022scgrec}, typically use the concatenation-based graph attention as follows:
\begin{equation}
    \alpha_{ij} = \frac{\exp(\text{MLP}[\mathbf e_{i},\mathbf e_{j}])}{\sum\nolimits_{n_k \in \mathcal{N}_{n_i}} \exp(\text{MLP}[\mathbf e_{i},\mathbf e_{j}])}.
\end{equation}

Also, other methods, including DICER~\cite{wu2019danser} and MEGCN~\cite{jin2020megcn}, use the similarity-based graph attention, which is another popular technique, \ie
\begin{equation}
    \alpha_{ij} = \frac{\exp(\text{sim}(\mathbf e_{i}, \mathbf e_{j}))}{\sum\nolimits_{n_k \in \mathcal{N}_{n_i}} \exp(\text{sim}(\mathbf e_{i}, \mathbf e_{k}))},
\end{equation}
where sim() denotes a similarity function such as cosine similarity and dot product. 

\subsubsection{\textbf{HetGNN}} The user-item interactions and user-user social relations can be regarded as the users' heterogeneous relationships, \ie a user's preferences on items and his/her friendship.
In this sense, a few methods~\cite{chen2021serec,wang2022dcan} have attempted to model the inputs as a heterogeneous graph and then design the HetGNN encoders for learning user and item embeddings, \ie
\begin{equation}
    \mathbf e_{i}^{(k)} = \sigma(\sum\nolimits_{n_j \in \mathcal{N}_{n_i}} \frac{1}{\sqrt{|\mathcal{N}_{n_j}||\mathcal{N}_{n_i}|}} \mathbf{e}_{j}^{(k-1)} \mathbf{W}_{v_{ij}}^{(k)}),
\end{equation}
where $v_{ij}$ indicates the type of relation between $n_i$ and $n_j$. 
As a result, the HetGNN encoder employs different transformation matrices according to the relations between two nodes.

For instance, SeRec~\cite{chen2021serec} defines four types of directed edges (\ie user-user edges, user-item edges, item-user edges, and item-item edges), constructing a heterogeneous graph based on the above edges. 
Then, it obtains each user $p_i$'s embedding $\mathbf{u}_i$ and each item $q_j$'s embedding $\mathbf{v}_j$ via the HetGNN encoder.

\subsubsection{\textbf{GRNN}} The sequential behaviors of users when they interact with items reflect the evolution of their preferences of items over time.
For this reason, time-aware recommender systems have attracted increasing attention in recent years~\cite{wang2021survey}.
Such temporal interactions are often divided into multiple user sessions and modeled as session-based \sorec. 
Multiple works~\cite{sun2020dgarecr,chen2022ssrgnn,lin2022gnndsr,liu2022gnnrec,niu2021mgsr,gu2021egfrec,song2019dgrec,wang2022mohcn,xiao2020mgnn} have attempted to model dynamic user interests through session-based or temporal \sorec.
These models leverage the GRNN encoders to capture these time-evolving interests. 

Suppose each user $p$ interacts with items in a given sequence $\mathcal{S}_p$. Consequently, one can create a sequence of interactions for each item $q$ as $\mathcal{S}_q$, consisting of users that rate item $q$ in a temporal sequence. In general, 
the temporal sequence is denoted for node $n_i$ as $\mathcal{S}_{n_i} = \{n^i_1, n^i_2, \cdots, n^i_K\}$.
Note that session-based encoders would divide $\mathcal{S}_{n_i}$ into multiple sessions $\mathcal{S}^t_{n_i}$ and encode each session separately. The GRNN encoder for node $n_i$ can be then generalized as:
\begin{equation}
    \mathbf e_i = \textsc{GRNN}(\mathcal{S}_{n_i}, \mathcal{N}_{n_i}),
\end{equation}
where \textsc{GRNN} is a combination of RNN and GNN modules. In particular, one can obtain dynamic user interests and item embeddings through a long short-term memory (LSTM)~\cite{peng2017cross,zayats2018conversation} unit, \ie
\begin{equation}
    \begin{aligned}
        \mathbf x_{i}^{(k)} &= \sigma(\mathbf{W}_x[\mathbf{h}_i^{(k-1)}, \mathbf n_k^i] + b_x), \\
        \mathbf f_{i}^{(k)} &= \sigma(\mathbf{W}_s[\mathbf{h}_i^{(k-1)}, \mathbf n_k^i] + b_s), \\
        \mathbf o_{i}^{(k)} &= \sigma(\mathbf{W}_o[\mathbf{h}_i^{(k-1)}, \mathbf n_k^i] + b_o), \\
        \mathbf{\tilde{c}}_{i}^{(k)} &= \tanh(\mathbf{W}_c[\mathbf{h}_j^{(k-1)}, \mathbf n_k^i] + b_c), \\
        \mathbf{c}_{i}^{(k)} &= \mathbf f_{i}^{(k)} \odot  \mathbf{c}_{i}^{(k-1)} + \mathbf x_{i}^{(k)} \odot \mathbf{\tilde{c}}_{i}^{(k)},\\
        \mathbf h_{i}^{(k)} &=  \mathbf o_{i}^{(k)} \odot \tanh(\mathbf{c}_{i}^{(k)}).
    \end{aligned}
\end{equation}

Then, the node embedding $\mathbf e_i$ is obtained using a GNN module such as GANN and GCN (as discussed below). In general, one can obtain
\begin{equation}
    \mathbf e_i = \textsc{GNN}(\mathbf{h}_i^{(K)}, \{\mathbf{h}_j^{(K)}:n_j\in\mathcal{N}_{n_i}\cup \{n_i\}\}).
\end{equation}

For instance, DREAM~\cite{song2020dream} obtains each user $p_i$'s embedding within each session using the GRNN encoder as above. It uses a Relational GAT module for the GNN layer to aggregate information from its social neighbors. Meanwhile, item embeddings $\mathbf v_j$ for item $q_j$ is obtained using a simple embedding layer.

\subsubsection{\textbf{HyperGNN}} Most GNN encoders, as mentioned above, learn pairwise connectivity between two nodes.
However, more-complicated connections can be captured by
jointly using user-item relations with user-user edges and/or using higher-order social relations.
For instance, triangular structures, including two users and their co-rated items, are a common motif.
To leverage such high-order relations, some works~\cite{yu2021mhcn,sun2022motifres,han2022dhhgcn} have attempted to model the inputs as a hypergraph and then design the HyperGNN encoders for learning user and item embeddings.

Let $\mathcal{G}=(\mathcal{N},\mathcal{E})$ denotes a hypergraph where $\mathcal{N}$ and $\mathcal{E}$ indicate sets of nodes and hyperedges, respectively. 
Each hyperedge $e \in \mathcal{E}$ is a subset of nodes, \textit{i.e.}, $e \in 2^{\mathcal{N}}$. The node degree is thus $d_i = \sum_{e \in\mathcal{E}: n_i \in e} {1} $ for $\forall n_i \in \mathcal{N}$. Then, each layer of the HyperGNN encoder learns node embeddings using relations as

\begin{equation}
    \mathbf h_i^{(k)} = \sigma(\sum\nolimits_{e \in \mathcal{E}: n_i \in e} \sum\nolimits_{n_j \in e} \frac{w_{e,i,j}}{\sqrt{d_i d_j} |e|} \mathbf h_{j}^{(k-1)}),
\end{equation}
where $w_{e,i,j}$ are learnable parameters and $d_i = \sum_{e \in\mathcal{E}: n_i \in e} {1}$. We note that HyperGNN-based \sorec methods~\cite{yu2021mhcn,han2022dhhgcn} remove non-linear activation and feature transformation as in the LightGCN encoder. 

For instance, MHCN~\cite{yu2021mhcn} designs three types of triangular motifs, constructing three incidence matrices, each representing a hypergraph induced by each motif.
Then, it obtains each user $p_i$'s embedding $\mathbf{u}_i$ via the multi-type HyperGNN encoders while obtaining each item $q_j$'s embedding $\mathbf{v}_j$ via the GCN encoder.

\subsubsection{\textbf{Others}} Furthermore, we briefly describe the two encoders, GAE and hyperbolic GNN, each of which is employed by only one method.
Liu et al.~\cite{liu2022siga} pointed out that GCN is mainly suitable for semi-supervised learning tasks.
On the other hand, they claimed that the goal of GAE coincides with that of the recommendation task, which is to minimize the reconstruction error of input and output~\cite{liu2022siga}. 
For this reason, they proposed a \sorec method, named SIGA, which employs GAE and is used for the rating prediction task. 

Meanwhile, Wang et al.~\cite{wang2021hypersorec} pointed out that since existing methods usually learn the user and item embeddings in the Euclidean space, these methods fail to explore the latent hierarchical property in the data. 
For this reason, they proposed a \sorec method, named HyperSoRec, which performs in the hyperbolic space because the exponential expansion of hyperbolic space helps preserve more-complex relationships between users and items~\cite{Krioukov2010}.

\begin{table*}
    \centering
  \caption{Taxonomy of decoder architectures}
  \label{tab:decoder}
    \resizebox{\textwidth}{!}{
  \renewcommand{\arraystretch}{1.7}
  \begin{tabular}{c|l}
    \toprule
    \textbf{Decoders} & \multicolumn{1}{c}{\textbf{Models}} \\
    \midrule
    \multirow{7}{*}{\textbf{Dot-product}} & DiffNet~\cite{wu2019diffnet}, DiffNet++~\cite{wu2020diffnet++}, DiffNetLG~\cite{song2021diffnetlg}, MEGCN~\cite{jin2020megcn}, ASR~\cite{jiang2021asr}, ATGCN~\cite{seng2021atgcn}, GNN-SOR~\cite{guo2020gnnsor}, \\
    & DGARec~\cite{sun2020dgarecr}, SAGLG~\cite{liu2021sagclg}, HIDM~\cite{li2020hidm}, SocialLGN~\cite{liao2022sociallgn}, GMAN~\cite{liao2022gman}, MPSR~\cite{liu2022mpsr}, DREAM~\cite{song2020dream}, \\
    &  SHGCN~\cite{zhu2021shgcn}, SCGRec~\cite{yang2022scgrec}, PDARec~\cite{zheng2021pdarec}, MHCN~\cite{yu2021mhcn}, SEPT~\cite{yu2021sept}, DcRec~\cite{wu2022dcrec}, \\
    &  DH-HGCN~\cite{han2022dhhgcn}, SSDRec~\cite{yan2022ssdrec}, SeRec~\cite{chen2021serec}, Motif-Res~\cite{sun2022motifres}, APTE~\cite{zhen2022apte}, EAGCN~\cite{wu2022eagcn}, \\
    & HOSR~\cite{liu2022hosr}, FeSoG~\cite{liu2022fesog}, IGRec~\cite{chen2022igrec}, ESRF~\cite{yu2022esrf}, 
    SDCRec~\cite{du2022sdcrec}, SoHRML~\cite{liu2022sohrml}, HyperSoRec~\cite{wang2021hypersorec}, DSR~\cite{sha2021dsr}\\
    &  DESIGN~\cite{tao2022design}, SRAN~\cite{xie2022sran}, IDiffNet~\cite{li2022idiffnet}, CGL~\cite{zhang2022cgl}, FBNE~\cite{chen2022fbne}, MrAPR~\cite{song2022mrapr}, GNNRec~\cite{liu2022gnnrec}, SGHAN~\cite{wei2022sghan}\\
    &   SSRGNN~\cite{chen2022ssrgnn}, DISGCN~\cite{li2022disgcn}, ME-LGN~\cite{miao2022melgn}, SGA~\cite{liufu2021sga}, SIGA~\cite{liu2022siga}, GDMSR~\cite{gdmsr23}, DGNN~\cite{dgnn23}, DSL~\cite{dsl23} \\ \midrule
    
    \multirow{6}{*}{\textbf{MLP}} & GraphRec~\cite{fan2019graphrec}, GraphRec+~\cite{fan2022graphrecp}, DICER~\cite{fu2021dicer}, SAN~\cite{jiang2021san}, KConvGraphRec~\cite{tien2020kconvgraphrec}, EGFRec~\cite{gu2021egfrec}, \\
    & FuseRec~\cite{narang2021fuserec}, GNNTSR~\cite{mandal2021gnntsr}, GAT-NSR~\cite{mu2019gatnsr}, TGRec~\cite{bai2020tgnn}, SoRecGAT~\cite{vijaikumar2019sorecgat}, PA-GAN~\cite{hou2021pagan}, \\ 
    & GHSCF~\cite{bi2021ghscf}, HeteroGraphRec~\cite{salamat2021heterographrec}, GLOW~\cite{leng2022glow}, BFHAN~\cite{zhao2021bfhan}, MGSR~\cite{niu2021mgsr}, GTN~\cite{hoang2021gtn}, \\
    & MGNN~\cite{xiao2020mgnn}, SR-HGNN~\cite{xu2020srhgnn}, DANSER~\cite{wu2019danser}, MutualRec~\cite{xiao2021mutualrec}, GSFR~\cite{xiao2022gsfr}, SOAP-VAE~\cite{walker2021soapvae}, \\
    & SENGR~\cite{shi2022sengr}, MOHCN~\cite{wang2022mohcn}, DCAN~\cite{wang2022dcan}, TAG~\cite{qiao2022tag}, HSGNN~\cite{wei2022hsgnn}, Social-RippleNet~\cite{jiang2022socialripplenet}, \\
    & GNN-DSR~\cite{lin2022gnndsr}, GDSRec~\cite{chen2022gdsrec}, DGRec~\cite{song2019dgrec}, DREAM~\cite{song2020dream} \\ 
    
    \bottomrule
  \end{tabular}
  }
\end{table*}

\subsection{Decoders}
In this subsection, we group the decoders of GNN-based \sorec into two categories: dot-product and multi-layer perceptron (MLP). 
Table~\ref{tab:decoder} summarizes the taxonomy of these decoders.

\begin{table*}
    \centering
  \caption{Taxonomy of loss functions}
  \label{tab:loss}
    \resizebox{\textwidth}{!}{
  \renewcommand{\arraystretch}{2.0}
  \renewcommand{\aboverulesep}{0.0pt}
    \renewcommand{\belowrulesep}{0.0pt}
    \renewcommand{\belowbottomsep}{1.722pt}
    \renewcommand{\abovetopsep}{2.798pt}
  \begin{tabular}{c|cc|l}
    \toprule
    \multicolumn{3}{c|}{\textbf{Loss Functions}} & \multicolumn{1}{c}{\textbf{Models}} \\
    \midrule
    \multirow{15}{*}{\begin{tabular}[c]{@{}c@{}} \textbf{Primary} \\ \textbf{Objectives} \end{tabular}}  & \multicolumn{2}{c|}{\multirow{5}{*}{\textbf{MSE}}} & GraphRec~\cite{fan2019graphrec},  GNNTSR~\cite{mandal2021gnntsr}, GAT-NSR~\cite{mu2019gatnsr}, TGRec~\cite{bai2020tgnn}, PA-GAN~\cite{hou2021pagan}, GHSCF~\cite{bi2021ghscf}, \\ 
    & & & GraphRec+~\cite{fan2022graphrecp}, GTN~\cite{hoang2021gtn}, PDARec~\cite{zheng2021pdarec}, 
    GNN-SOR~\cite{guo2020gnnsor}, DANSER~\cite{wu2019danser}, SAN~\cite{jiang2021san}, \\
    & & & KConvGraphRec~\cite{tien2020kconvgraphrec},  HeteroGraphRec~\cite{salamat2021heterographrec}, GMAN~\cite{liao2022gman},  SR-HGNN~\cite{xu2020srhgnn}, DGARec-R~\cite{sun2020dgarecr}, \\ 
    & & &  MGSR~\cite{niu2021mgsr}, APTE~\cite{zhen2022apte}, EAGCN~\cite{wu2022eagcn}, FeSoG~\cite{liu2022fesog}, SENGR~\cite{shi2022sengr}, MOHCN~\cite{wang2022mohcn}, TAG~\cite{qiao2022tag},\\
    & & & Social-RippleNet~\cite{jiang2022socialripplenet}, GNN-DSR~\cite{lin2022gnndsr}, GDSRec~\cite{chen2022gdsrec} \\ \cmidrule{2-4}

    & \multicolumn{2}{c|}{\multirow{5}{*}{\textbf{BPR}}} & ASR~\cite{jiang2021asr}, SAGLG~\cite{liu2021sagclg}, MGNN~\cite{xiao2020mgnn}, HIDM~\cite{li2020hidm}, SocialLGN~\cite{liao2022sociallgn}, MPSR~\cite{liu2022mpsr}, SHGCN~\cite{zhu2021shgcn},  \\
    & & & MutualRec~\cite{xiao2021mutualrec}, ATGCN~\cite{seng2021atgcn}, DiffNet~\cite{wu2019diffnet}, DiffNet++~\cite{wu2020diffnet++}, DiffNetLG~\cite{song2021diffnetlg},  MEGCN~\cite{jin2020megcn}, \\ 
    & & &  GLOW~\cite{leng2022glow}, SCGRec~\cite{yang2022scgrec}, SEPT~\cite{yu2021sept}, DcRec~\cite{wu2022dcrec}, MHCN~\cite{yu2021mhcn}, DH-HGCN~\cite{han2022dhhgcn}, GSFR~\cite{xiao2022gsfr}, HOSR~\cite{liu2022hosr}  \\
    & & &  IGRec~\cite{chen2022igrec}, ESRF~\cite{yu2022esrf}, SoHRML~\cite{liu2022sohrml}, DSR~\cite{sha2021dsr}, SRAN~\cite{xie2022sran}, IDiffNet~\cite{li2022idiffnet}, CGL~\cite{zhang2022cgl}, MrAPR~\cite{song2022mrapr}, DISGCN~\cite{li2022disgcn}\\
    & & &    HSGNN~\cite{wei2022hsgnn}, ME-LGN~\cite{miao2022melgn}, SGA~\cite{liufu2021sga}, Motif-Res~\cite{sun2022motifres}, GDMSR~\cite{gdmsr23}, MADM~\cite{madm24}, DGNN~\cite{dgnn23}, DSL~\cite{dsl23} \\ \cmidrule{2-4}
    
    & \multicolumn{2}{c|}{\multirow{3}{*}{\textbf{CE}}} & DICER~\cite{fu2021dicer}, SoRecGAT~\cite{vijaikumar2019sorecgat}, DANSER~\cite{wu2019danser}, BFHAN~\cite{zhao2021bfhan}, EGFRec~\cite{gu2021egfrec}, FuseRec~\cite{narang2021fuserec},  \\
    & & & SeRec~\cite{chen2021serec}, SSDRec~\cite{yan2022ssdrec},
    DESIGN~\cite{tao2022design}, FBNE~\cite{chen2022fbne}, GNNRec~\cite{liu2022gnnrec}, DCAN~\cite{wang2022dcan}, SOAP-VAE~\cite{walker2021soapvae} \\
    & & & SGHAN~\cite{wei2022sghan}, SSRGNN~\cite{chen2022ssrgnn}, GDSRec~\cite{chen2022gdsrec}, 
    SIGA~\cite{liu2022siga}, DGRec~\cite{song2019dgrec}, DREAM~\cite{song2020dream} \\ \cmidrule{2-4}
    
    & \multicolumn{2}{c|}{\textbf{Hinge}} & HyperSoRec~\cite{wang2021hypersorec} \\ \midrule
    
    \multirow{12}{*}{\begin{tabular}[c]{@{}c@{}} \textbf{Auxiliary} \\ \textbf{Objectives} \end{tabular}}  
    & \multicolumn{2}{c|}{\textbf{Social LP}} & MGNN~\cite{xiao2020mgnn}, MutualRec~\cite{xiao2021mutualrec}, SR-HGNN~\cite{xu2020srhgnn}, APTE~\cite{zhen2022apte}, SoHRML~\cite{liu2022sohrml}, FBNE~\cite{chen2022fbne}, GDMSR~\cite{gdmsr23}  \\ \cmidrule{2-4}
    
    & \multirow{3}{*}{\textbf{SSL}} & \textbf{Social} & SEPT~\cite{yu2021sept}, DcRec~\cite{wu2022dcrec}, MADM~\cite{madm24}, DSL~\cite{dsl23}\\    
    & & \textbf{Interaction} & SDCRec~\cite{du2022sdcrec}, CGL~\cite{zhang2022cgl}, DISGCN~\cite{li2022disgcn}, DCAN~\cite{wang2022dcan}, DcRec~\cite{wu2022dcrec} \\
    & & \textbf{Motif} & Motif-Res~\cite{sun2022motifres}, MHCN~\cite{yu2021mhcn}  \\ \cmidrule{2-4}
    & \multicolumn{2}{c|}{\textbf{Group}} & GLOW~\cite{leng2022glow}, GMAN~\cite{liao2022gman} \\ \cmidrule{2-4}
    & \multicolumn{2}{c|}{\textbf{Adv}} & ESRF~\cite{yu2022esrf} \\ \cmidrule{2-4}
    & \multicolumn{2}{c|}{\textbf{Path}} & SPEX~\cite{li2021spex} \\ \cmidrule{2-4}
    & \multicolumn{2}{c|}{\textbf{KD}} & DESIGN~\cite{tao2022design} \\ \cmidrule{2-4}
    & \multicolumn{2}{c|}{\textbf{Sentiment}} & SENGR~\cite{shi2022sengr} \\ \cmidrule{2-4}
    & \multicolumn{2}{c|}{\textbf{Policy Net.}} & DANSER~\cite{wu2019danser} \\
    \bottomrule
  \end{tabular}
  }
\end{table*}

\subsubsection{\textbf{Dot-product}} Many methods~\cite{wu2019diffnet,wu2020diffnet++,song2021diffnetlg,jin2020megcn,jiang2021asr,seng2021atgcn,guo2020gnnsor,sun2020dgarecr,liu2021sagclg,li2020hidm,liao2022sociallgn,liao2022gman,liu2022mpsr,song2020dream,zhu2021shgcn,yang2022scgrec,zheng2021pdarec,yu2021mhcn,yu2021sept,wu2022dcrec,han2022dhhgcn,yan2022ssdrec,chen2021serec,sun2022motifres,zhen2022apte,wu2022eagcn,liu2022hosr,liu2022fesog,chen2022igrec,yu2022esrf,zhu2022sinews,du2022sdcrec,liu2022sohrml,wang2021hypersorec,sha2021dsr,tao2022design,xie2022sran,li2022idiffnet,zhang2022cgl,chen2022fbne,song2022mrapr,liu2022gnnrec,wei2022sghan,chen2022ssrgnn,li2022disgcn,miao2022melgn,liufu2021sga,liu2022siga} simply predict a user $p_i$'s preference $\hat{r}_{ij}$ on an item $q_j$ via a dot product of their corresponding embeddings, \ie
\begin{equation}
    \hat{r}_{ij} = \mathbf{u}_i \cdot \mathbf{v}_j^\top.
\end{equation}

\subsubsection{\textbf{MLP}} More than half of the existing methods~\cite{fan2019graphrec,fan2022graphrecp,fu2021dicer,jiang2021san,tien2020kconvgraphrec,gu2021egfrec,narang2021fuserec,mandal2021gnntsr,mu2019gatnsr,bai2020tgnn,vijaikumar2019sorecgat,hou2021pagan,bi2021ghscf,salamat2021heterographrec,leng2022glow,zhao2021bfhan,niu2021mgsr,hoang2021gtn,xiao2020mgnn,xu2020srhgnn,wu2019danser,xiao2021mutualrec,xiao2022gsfr,walker2021soapvae,shi2022sengr,wang2022mohcn,wang2022dcan,qiao2022tag,wei2022hsgnn,jiang2022socialripplenet,lin2022gnndsr,chen2022gdsrec,song2019dgrec,song2020dream} predict a user $p_i$'s preference $\hat{r}_{ij}$ on an item $q_j$ by employing MLP as follows:
\begin{equation}
    \hat{r}_{ij} = \sigma_{L}(\mathbf{W}^\top_{L}(\sigma_{L-1}(...\sigma_{2}(\mathbf{W}_{2}^\top \begin{bmatrix}
        \mathbf{u}_{i} \\
        \mathbf{v}_{j}
    \end{bmatrix}
    +\mathbf{b}_{2})...))+\mathbf{b}_{L}, \\
\end{equation}
where $\mathbf{W}_{i}$, $\mathbf{b}_{i}$, and $\sigma_{i}$ denote the weight matrix, bias vector, and activation function for $i$-th layer's perceptron, respectively.

\subsection{Loss Functions}
In this subsection, we first group the primary loss functions of GNN-based \sorec into 4 categories: Bayesian personalized ranking (BPR)~\cite{RendleFGS09}, mean squared error (MSE), cross-entropy (CE), and hinge loss. 
In addition, we found that some works additionally employ auxiliary loss functions. 
Thus, we further group these loss functions into 8 categories: social link prediction (LP) loss, self-supervised loss (SSL), group-based loss, adversarial (Adv) loss, path-based loss, knowledge distillation (KD) loss, sentiment-aware loss, and policy-network-based (Policy Net) loss.
Table~\ref{tab:loss} summarizes the taxonomy of loss functions used in existing work.

\subsubsection{\textbf{Primary Loss Functions}} Different primary loss functions are employed depending on whether the methods focus on explicit or implicit feedback. 

\vspace{1mm}
\textbf{MSE Loss.} For the methods that focus on explicit feedback (\eg star ratings) of users, most of them~\cite{fan2019graphrec,mandal2021gnntsr,mu2019gatnsr,bai2020tgnn,hou2021pagan,bi2021ghscf,fan2022graphrecp,hoang2021gtn,zheng2021pdarec,guo2020gnnsor,wu2019danser,jiang2021san,tien2020kconvgraphrec,salamat2021heterographrec,xu2020srhgnn,liao2022gman,sun2020dgarecr,niu2021mgsr,zhen2022apte,wu2022eagcn,liu2022fesog,shi2022sengr,wang2022mohcn,qiao2022tag,jiang2022socialripplenet,lin2022gnndsr,chen2022gdsrec} learn user and item embeddings via the MSE-based loss function $\mathcal{L}_{MSE}$, which is defined as follows:
\begin{equation}
\mathcal{L}_{MSE}=\sum_{p_i\in\mathcal{U}}\sum_{q_j\in\mathcal{I}}(\hat{r}_{ij}-{r}_{ij})^2,
\end{equation}
where ${r}_{ij}$ indicates $p_i$'s real rating score on $q_j$. That is, the embeddings of $p_i$ and $q_j$ are learned, aiming at minimizing the differences between $p_i$'s real and predicted scores, \ie ${r}_{ij}$ and $ \hat{r}_{ij}$, for $q_j$.

\vspace{1mm}
\textbf{BPR Loss.} For the methods that focus on implicit feedback (\eg click or browsing history) of users, most of them~\cite{jiang2021asr,liu2021sagclg,xiao2020mgnn,li2020hidm,liao2022sociallgn,liu2022mpsr,zhu2021shgcn,xiao2021mutualrec,seng2021atgcn,wu2019diffnet,wu2020diffnet++,song2021diffnetlg,jin2020megcn,leng2022glow,yang2022scgrec,yu2021sept,wu2022dcrec,yu2021mhcn,han2022dhhgcn,xiao2022gsfr,liu2022hosr,chen2022igrec,yu2022esrf,liu2022sohrml,sha2021dsr,xie2022sran,li2022idiffnet,zhang2022cgl,song2022mrapr,li2022disgcn,wei2022hsgnn,miao2022melgn,liufu2021sga,sun2022motifres} learn user and item embeddings via the BPR-based loss function $\mathcal{L}_{BPR}$, which is defined as follows:
\begin{equation}
\mathcal{L}_{BPR}=-\sum_{p_i\in\mathcal{U}}\sum_{q_j\in\mathcal{N}_{p_i}}\sum_{q_k\in\mathcal{I}\text{\textbackslash}\mathcal{N}_{p_i}}\text{log}\sigma(\hat{r}_{ij}-\hat{r}_{ik}),
\end{equation}
where $\mathcal{U}$ and $\mathcal{N}_{n_i}$ denote a set of users and a set of items rated by $p_i$, respectively. $\hat{r}_{ij}$ and $\hat{r}_{ik}$ indicate $p_i$'s preference on the rated item $q_j$ and the (randomly-sampled) unrated item $q_k$, respectively.
Also, $\sigma$ indicates the sigmoid function.
That is, the embeddings of $p_i$, $q_j$, and $q_k$ are learned based on the intuition that $p_i$’s preference $\hat{r}_{ij}$ on $q_j$ is likely to be higher than $p_i$’s preference $\hat{r}_{ik}$ on $q_k$.

\vspace{1mm}
\textbf{CE Loss.} Several methods~\cite{fu2021dicer,vijaikumar2019sorecgat,wu2019danser,zhao2021bfhan,gu2021egfrec,narang2021fuserec,chen2021serec,yan2022ssdrec,zhu2022sinews,tao2022design,chen2022fbne,liu2022gnnrec,wang2022dcan,walker2021soapvae,wei2022sghan,chen2022ssrgnn,chen2022gdsrec,li2021spex,liu2022siga,song2019dgrec,song2020dream} for implicit feedback learn user and item embeddings via the CE-based loss function $\mathcal{L}_{CE}$, which is defined as follows:
\begin{equation}
\mathcal{L}_{CE}=-\sum_{p_i\in\mathcal{U}}\sum_{q_j\in\mathcal{I}}{r}_{ij}\text{log}(\hat{r}_{ij}) + (1-{r}_{ij})\text{log}(1-\hat{r}_{ij}),
\end{equation}
where $\mathcal{I}$ indicates a set of items. It should be noted that ${r}_{ij}=1$ if $q_j \in \mathcal{N}_{p_i}$, otherwise ${r}_{ij}=0$. 
That is, the embeddings of $p_i$ and $q_j$ are learned, aiming at maximizing $p_i$’s preferences on his/her rated items while minimizing $p_i$'s preferences on his/her unrated items.

\vspace{1mm}
\textbf{Hinge Loss.} A method~\cite{wang2021hypersorec} for implicit feedback learns user and item embeddings via the hinge loss function $\mathcal{L}_{Hinge}$, which is defined as follows:
\begin{equation}
\mathcal{L}_{Hinge}=\sum_{p_i\in\mathcal{U}}\sum_{q_j\in\mathcal{N}_{p_i}}\sum_{q_k\in\mathcal{I}\text{\textbackslash}\mathcal{N}_{p_i}}\text{max}(0,\lambda+(\hat{r}_{ij})^2-(\hat{r}_{ik})^2),
\end{equation}
where $\lambda$ indicates the safety margin size. That is, the embeddings of $p_i$, $q_j$, and $q_k$ are learned, aiming at ensuring that $p_i$’s preferences on his/her rated items $q_j$ are higher than those on his/her unrated items $q_k$ at least by a margin of $\lambda$.

\subsubsection{\textbf{Auxiliary Loss Functions}} 
Here, we discuss the auxiliary loss functions used by GNN-based \sorec methods.

\vspace{1mm}
\textbf{Social Link Prediction (LP) Loss.} 
It should be noted that the primary objectives of the existing works focus on reconstructing the input U-I rating graph.
Along with this, papers like MGNN~\cite{xiao2020mgnn}, MutualRec~\cite{xiao2021mutualrec}, and SR-HGNN~\cite{xu2020srhgnn} learn the BPR-based social LP loss that aims at reconstructing the input U-U social graph.
Through this method, user embeddings can be informed further to reconstruct the social relations, which allows them to better capture the social network structure that is essential for the more-effective social recommendation.

\vspace{1mm}
\textbf{Self-Supervised Loss (SSL)}. SSL originated in image and text domains to address the deficiency of labeled data~\cite{liu2020selfsupervised}. 
The basic idea of SSL is to assign labels for unlabeled data and exploit them additionally in the training process.
It is well-known that the data sparsity problem significantly affects the performance of recommender systems.
Therefore, there has recently been a surge of interest in SSL for recommender systems~\cite{yu2022ssl}.

Some GNN-based \sorec methods~\cite{yu2021sept,yu2021mhcn,wu2022dcrec,zhang2022cgl,li2022disgcn,wang2022dcan,du2022sdcrec,wu2022dcrec,sun2022motifres} designed SSL, which is derived from U-U social and/or U-I rating graphs.
In this survey, we categorized them as social SSL and interaction-based SSL depending on the graph type employed to design SSL.
For the social SSL, SEPT~\cite{yu2021sept} augments different views related to users with the U-U social graph and designs two socially-aware encoders that aim at reconstructing the augmented views. It adopts the regime of tri-training~\cite{ZhouL05tri}, which operates on the augmented views above for self-supervised signals.
For the interaction-based SSL, SDCRec~\cite{du2022sdcrec} samples two items among items rated by a user, which have the highest similarities to the user. Then, it additionally utilizes them as self-supervised signals.

On the other hand, Motif-Res~\cite{sun2022motifres} and MHCN~\cite{yu2021mhcn} explore the motif information in graph structure so that such information can be utilized as self-supervised signals.
For instance, MHCN~\cite{yu2021mhcn} constructs multi-type hyperedges, which are instances of a set of triangular relations, and designs SSL by leveraging the hierarchy in the hypergraph structures. It aims at reflecting the user node’s local and global high-order connectivity patterns in different hypergraphs~\cite{yu2021mhcn}. 

\vspace{1mm}
\textbf{Group-based Loss.} GLOW~\cite{leng2022glow} and GMAN~\cite{liao2022gman} make use of the user groups. 
Based on the group information, both methods additionally design the group-based loss. 
They define the group-item interaction as indicating a set of users that have interacted with an item.
Then, they represent each group's embedding by attentively aggregating the users' embeddings within the corresponding group.
Finally, the user and item embeddings are learned via a group-based loss so that each group's preferences on items rated by users in the corresponding group are likely to be higher than those of their unrated items.

\vspace{1mm}
\textbf{Others.} We briefly discuss the other loss functions that are employed by only one method.
Yu et al.~\cite{yu2022esrf} designed an adversarial mechanism to consider the fact that social relations are very sparse, noisy, and multi-faceted in real-world social networks.
On the other hand, Li et al.~\cite{li2021spex} pointed out that existing \sorec methods fail to distinguish social influence from social homophily. To address this limitation, they designed an auxiliary loss function that models and captures the rich information conveyed by the formation of social homophily~\cite{li2021spex}. 
Furthermore, Tao et al.~\cite{tao2022design} leveraged the knowledge distillation (KD) technique into the social recommendation to address the overfitting problem of existing methods.
Shi et al.~\cite{shi2022sengr} incorporated both sentiment information derived from reviews and interaction information captured by the GNN encoder. To this end, they designed an auxiliary loss function that captures different sentimental aspects of items from reviews~\cite{shi2022sengr}.
Lastly, Wu et al.~\cite{wu2019danser} designed a policy-based loss function based on a contextual multi-armed bandit~\cite{BubeckC12}, which dynamically weighs different social effects, \ie social homophily, social influence, item-to-item homophily, and item-to-item influence.

\begin{table}[t]
    \centering
    \caption{Comparison of time complexity for GNN-based SocialRS methods. It should be noted that only methods that discuss their complexity in the respective papers are listed.}
    \label{tab:complexity}
    \resizebox{\textwidth}{!}{
    \renewcommand{\arraystretch}{1.2}
    \renewcommand{\aboverulesep}{0.0pt}
    \renewcommand{\belowrulesep}{0.0pt}
    \renewcommand{\belowbottomsep}{1.722pt}
    \renewcommand{\abovetopsep}{2.798pt}
    \newcolumntype{H}{>{\setbox0=\hbox\bgroup}c<{\egroup}@{}}
    \begin{tabular}{c|c|c|c|c}
        \toprule
        \multicolumn{3}{c|}{\textbf{Encoders}} & \multirow{2}{*}{\textbf{Models}} & \multirow{2}{*}{\textbf{Time Complexity}} \\ \cmidrule{1-3}
        \textbf{User Social} & \textbf{User Interest} & \textbf{Item Encoder} &  &  \\
        \midrule 
        \multirow{14}{*}{GANN} & \multirow{12}{*}{GANN} & \multirow{8}{*}{GANN} &  DiffNet++~\cite{wu2020diffnet++} & $O(m(L_s+L_i)D+nL_uD)$ \\
        & & & SR-HGNN~\cite{xu2020srhgnn} & $O(|\mathbf{R}|d)$ \\
        & & & DISGCN~\cite{li2022disgcn} & $O(|B|d^2K+(|B|+|\mathbf{S}^{+}|+|\mathbf{R}|)dK+|\mathbf{R}|d))$ \\
        & & & ME-LGN~\cite{miao2022melgn} & $O(mKs)$ \\
        & & & GDSRec~\cite{chen2022gdsrec} & $O(((m+n)s+<Q>)D)$ \\
        & & & GraphRec+~\cite{fan2022graphrecp} & $O(m(L_i+L_s)d+n(L_u+m_c)d)$ \\
        & & & ESRF~\cite{yu2022esrf} & $O(|\mathbf{R}|d+|\mathbf{S}|d+kmd)$ \\
        & & & SoHRML~\cite{liu2022sohrml} & $O(|\mathbf{R}|+|\mathbf{S}|)(2Kd_1d_2+\sum_k=1 d_kd_{k-1}))$ \\ \cmidrule{3-5}
        &  & \multirow{2}{*}{Emb} &  SAN~\cite{jiang2021san} & $O(mL_sK)$ \\
        & & & HSGNN~\cite{wei2022hsgnn} & $O(mKs)$ \\ \cmidrule{3-5}
        &  & \multirow{2}{*}{GCN} & BFHAN~\cite{zhao2021bfhan} & $O(K(|\mathbf{R}|d+(m+n)(d2+d)))$ \\
        & & & SHGCN~\cite{zhu2021shgcn} & $O((|\mathbf{S}|+|\mathbf{E}|)d)$ \\ \cmidrule{2-5}
        & GRNN & GRNN & GNN-DSR~\cite{lin2022gnndsr} & $O(m(2L_i+L_s)d^2 + n(2L_u+m_c)d^2)$ \\ \cmidrule{2-5}
         & \multirow{1}{*}{RNN} & RNN  & SGHAN~\cite{wei2022sghan} & $O(hMd)+O(hdK)$ \\ \midrule
        \multirow{4}{*}{GCN} & \multirow{3}{*}{GCN} & \multirow{3}{*}{Emb} & DiffNet~\cite{wu2019diffnet} & $O(mKL_s)$\\
        & & & MEGCN~\cite{jin2020megcn} & $O(mk_1k_2)$ \\ 
        & & & HOSR~\cite{liu2022hosr} & $O(K|\mathbf{S}|d^2+|\mathbf{R}|d)$ \\ \cmidrule{2-5}
        & MLP & Emb & MGNN~\cite{xiao2020mgnn} & $O(m^2Kd^2)$ \\ \midrule
          
         \multirow{6}{*}{LightGCN} & \multirow{6}{*}{LightGCN} & \multirow{5}{*}{LightGCN} & SocialLGN~\cite{liao2022sociallgn} & $O(m(L_s+L_i+d)d+nL_ud)$ \\
         & & & EAGCN~\cite{wu2022eagcn} & $O(m+nd^2+(|\mathbf{R}|+|\mathbf{S}|)dK+|\mathbf{R}|d)$ \\
         & & & CGL~\cite{zhang2022cgl} & $O(|\mathbf{R}|d(K+1)+|\mathbf{S}|dK+5Bd+B^2d)$ \\
         & & & SEPT~\cite{yu2021sept} & $O(|\mathbf{R}|d+mlog(K))$ \\
         & & & DSR~\cite{sha2021dsr} & $O((L_S|\mathbf{S}|td + L_R|\mathbf{R}|d)+mfd^2)$\\ \cmidrule{3-5}
         & & Emb &  IDiffNet~\cite{li2022idiffnet} & $O(mKL_i+nKL_u)$\\ \midrule
        
        \multirow{1}{*}{HyperGNN} & \multirow{1}{*}{GANN} & \multirow{1}{*}{HyperGNN} &  MHCN~\cite{yu2021mhcn} & $O(|\mathbf{A}+|dK)$ \\ \midrule        
        Hyperbolic & Hyperbolic & Hyperbolic & HyperSoRec~\cite{wang2021hypersorec} & $O(cb\prod_{i=1}^K |N_i|)$ \\
        
        \bottomrule
    \end{tabular}}
\end{table}

\subsection{Model Complexity}
Finally, we conduct a time complexity analysis for GNN-based SocialRS methods. In Table~\ref{tab:complexity}, we present a summary of the time complexity of the methods, providing values from the corresponding papers. The common notations used for the time complexity are outlined below.
\begin{itemize}
    \item $m$ and $n$: the number of users and items, respectively;
    \item $|\mathbf{R}|$, $|\mathbf{S}|$, and $|\mathbf{E}|$: number of edges in the user-item interaction, user-user social, and hypergraphs, respectively;
    \item $|\mathbf{S}^+|$: number of all the friend pairs with social influence;
    \item $K$ and $d$: number of GNN layers and the embedding size, respectively;
    \item $L_S$ and $L_R$: number of layers for social and rating graphs, respectively;
    \item $L_s$ and $L_i$: average number of social and item neighbors per user, respectively;
    \item $L_u$ and $m_c$: average number of user and item neighbors per item, respectively;
    \item $s$ and $km$: number of the sampled neighbors and alternative neighbors, respectively;
    \item $t$, $h$, and $B$: number of iterations, number of LSTM hidden state, and the batch size, respectively.
\end{itemize}
For the remaining model-specific notations, including $M$, $k_1$, $k_2$, $f$, $c$, and $b$, please refer to the corresponding papers. From Table~\ref{tab:complexity}, we observe that methods using encoders such as RNN and HetGNN, which have high complexity, do not offer detailed insights into their computational demands. 
On the contrary, most methods using LightGCN discuss their complexity and also substantiate their efficiency, including scalability, through experimental validation.

\begin{table*}[t]
    \centering
    \caption{Statistics of 17 publicly-available benchmark datasets. Dataset source is hyperlinked to each dataset name: datasets colored {\color{blue}blue} contain links for both user-item interactions and user-user relations, whereas the ones in {\color{red}red} only contain one of the two due to unavailability of the other.} \label{tab:datasets}
    \resizebox{\textwidth}{!}{
    \begin{threeparttable}
    \renewcommand{\arraystretch}{2.0}
    \renewcommand{\aboverulesep}{0.0pt}
    \renewcommand{\belowrulesep}{0.0pt}
    \renewcommand{\belowbottomsep}{1.722pt}
    \renewcommand{\abovetopsep}{2.798pt}
    \begin{tabular}{c|c|cc|cc|l}
    \toprule
    \textbf{Domains} & \textbf{Datasets} & \textbf{\# Users} & \textbf{\# Items} & \textbf{\# Ratings} & \textbf{\# Social} & 
    \multicolumn{1}{c}{\textbf{Papers Used}} \\
    \midrule
    \multirow{8}{*}{\textbf{Product}} & \multirow{3}{*}{\href{https://www.cse.msu.edu/~tangjili/trust.html}{\color{blue}\textbf{Epinions}}} & \multirow{3}{*}{18,088} & \multirow{3}{*}{261,649} & \multirow{3}{*}{764,352} & \multirow{3}{*}{355,813} & \cite{zheng2021pdarec,guo2020gnnsor,xiao2021mutualrec,fan2019graphrec,fu2021dicer,mandal2021gnntsr,mu2019gatnsr,bai2020tgnn,hou2021pagan,xiao2020mgnn,bi2021ghscf,li2020hidm,walker2021soapvae,fan2022graphrecp,hoang2021gtn},\\ 
    & & & & & &  \cite{wu2020diffnet++, wu2019danser,tien2020kconvgraphrec,salamat2021heterographrec,xu2020srhgnn, zhao2021bfhan,narang2021fuserec,sun2020dgarecr,xiao2022gsfr,zhen2022apte,liu2022fesog,du2022sdcrec,liu2022sohrml}, \\ 
    & & & & & & \cite{lin2022gnndsr,chen2022gdsrec,li2021spex,song2020dream,sha2021dsr,tao2022design,wang2021hypersorec,wang2022mohcn,liu2022gnnrec} \\  \cmidrule{2-7}
    & \multirow{3}{*}{\href{http://www.cse.msu.edu/~tangjili/trust.html}{\color{blue}\textbf{Ciao}}} & \multirow{3}{*}{7,317} & \multirow{3}{*}{104,975} & \multirow{3}{*}{283,319} & \multirow{3}{*}{111,781} & \cite{mandal2021gnntsr,bai2020tgnn,hou2021pagan,bi2021ghscf,li2020hidm,liao2022sociallgn,walker2021soapvae,liu2022mpsr,tien2020kconvgraphrec,fan2022graphrecp,salamat2021heterographrec,xu2020srhgnn,zhao2021bfhan,narang2021fuserec,sun2020dgarecr},  \\ 
    & & & & & & \cite{wang2021hypersorec,fan2019graphrec,fu2021dicer,jiang2021asr,hoang2021gtn,wu2022dcrec,liu2022siga,xiao2022gsfr,wu2022eagcn,liu2022fesog,du2022sdcrec,liu2022sohrml,sha2021dsr,tao2022design,chen2022gdsrec},\\
    & & & & & & \cite{wang2022mohcn,song2022mrapr,lin2022gnndsr} \\ \cmidrule{2-7}
    & \href{https://github.com/tsinghua-fib-lab/DISGCN}{\color{blue}{\textbf{Beidan}}}
    & 2,841 & 2,298 & 35,146 & 2,367 & \cite{li2022disgcn} \\ \cmidrule{2-7}
    & \href{https://github.com/tsinghua-fib-lab/DISGCN}{\color{blue}{\textbf{Beibei}}} & 24,827 & 16,864 & 1,667,320 & 197,590 & \cite{li2022disgcn} \\ \midrule
    \multirow{5}{*}{\textbf{Location}} & \multirow{2}{*}{\href{https://github.com/librahu/HIN-Datasets-for-Recommendation-and-Network-Embedding}{\color{blue}\textbf{Yelp}}\footnotemark} & \multirow{2}{*}{19,539} & \multirow{2}{*}{21,266} & \multirow{2}{*}{405,884} & \multirow{2}{*}{363,672} & \cite{jiang2021asr,vijaikumar2019sorecgat,liu2022mpsr,guo2020gnnsor,wu2019diffnet,wu2020diffnet++,song2021diffnetlg,jin2020megcn,jiang2021san,yu2021sept,yu2021mhcn,han2022dhhgcn,sun2022motifres,zhen2022apte}, \\
    & & & & & & \cite{wu2022eagcn,liu2022hosr,tao2022design,wang2021hypersorec,xie2022sran,li2022idiffnet,zhang2022cgl,wang2022mohcn,song2022mrapr,wei2022sghan,miao2022melgn,song2019dgrec,shi2022sengr,chen2022fbne,qiao2022tag}\\ \cmidrule{2-7}
    & \href{https://lihui.info/data/dianping/}{\color{blue}\textbf{Dianping}} & 59,426 & 10,224 & 934,334 & 813,331 & \cite{wu2020diffnet++,wu2022dcrec} \\  \cmidrule{2-7}
    & \href{https://snap.stanford.edu/data/loc-gowalla.html}{\color{blue}\textbf{Gowalla}} & 33,661 & 41,229 & 1,218,599 & 283,778 & \cite{seng2021atgcn,chen2021serec,li2022atstggnn,wu2022eagcn,yu2022esrf,wang2022dcan,wei2022sghan,chen2022ssrgnn,liufu2021sga} \\ \cmidrule{2-7}
    & \href{https://sites.google.com/site/yangdingqi/home/foursquare-dataset}{\color{blue}\textbf{Foursquare}} & 39,302 & 45,595 & 3,627,093 & 304,030 & \cite{chen2021serec,li2022atstggnn,wang2022dcan,chen2022ssrgnn} \\ \midrule
    \multirow{3}{*}{\textbf{Movie}} & \href{https://grouplens.org/datasets/movielens/}{\color{red}{\textbf{MovieLens}}}  & 138,159 & 16,954 & 1,501,622 & 487,184 & \cite{liu2021sagclg,tien2020kconvgraphrec,jiang2022socialripplenet,chen2022fbne} \\ \cmidrule{2-7}
    & \href{http://datasets.syr.edu/datasets/Flixster.html}{\color{red}\textbf{Flixster}} &  58,470 &  38,076 & 3,619,736 & 667,313 & \cite{xiao2020mgnn,fan2022graphrecp,guo2020gnnsor,xiao2021mutualrec,xiao2022gsfr,liu2022sohrml,liu2022siga}\\ \cmidrule{2-7}
    & \href{https://github.com/npxiaoying/Social-Recommendation/tree/master/dataset/Filmtrust}{\color{blue}\textbf{FilmTrust}} & 1,508 & 2,071 & 35,497 & 1,853 & \cite{jiang2021asr,mu2019gatnsr,zheng2021pdarec,sun2022motifres,liu2022fesog,liu2022siga} \\ \midrule
    \multirow{1}{*}{\textbf{Image}} & \href{http://datasets.syr.edu/datasets/Flickr.html}{\color{red}\textbf{Flickr}} & 8,358 & 82,120 & 327,815 & 187,273 & \cite{wu2019diffnet,wu2020diffnet++,jin2020megcn,jiang2021san,wu2022eagcn,tao2022design,xie2022sran,li2022idiffnet,zhang2022cgl} \\ \midrule
    \multirow{1}{*}{\textbf{Music}} & \href{https://files.grouplens.org/datasets/hetrec2011/hetrec2011-lastfm-2k.zip}{\color{blue}{\textbf{Last.fm}}} & 1,892 & 17,632 & 92,834 & 25,434 & \cite{liao2022sociallgn, xiao2021mutualrec, seng2021atgcn,tien2020kconvgraphrec,yu2021sept,yu2021mhcn,zhang2021kcrec,chen2022igrec,yu2022esrf,liufu2021sga,miao2022melgn} \\ \midrule
    \multirow{1}{*}{\textbf{Bookmark}} & \href{https://files.grouplens.org/datasets/hetrec2011/hetrec2011-delicious-2k.zip}{\color{blue}{\textbf{Delicious}}} & 1,629 & 3,450 & 282,482 & 12,571 & \cite{li2020hidm,gu2021egfrec, chen2021serec,wang2022dcan,chen2022ssrgnn,lin2022gnndsr,song2019dgrec} \\  \midrule
    \multirow{2}{*}{\textbf{Microblog}} & \href{https://www.aminer.cn/data-sna\#Weibo-Net-Tweet}{\color{blue}{\textbf{Weibo}}} & 6,812 & 19,519 & 157,555 & 133,712 & \cite{li2021spex} \\ \cmidrule{2-7}
    & \href{https://www.aminer.cn/data-sna\#Twitter-Dynamic-Net}{\color{blue}{\textbf{Twitter}}} & 8,930 & 232,849 & 466,259 & 96,718 & \cite{li2021spex} \\ \midrule
    \multirow{2}{*}{\textbf{Miscellaneous}} & \multirow{2}{*}{\href{https://github.com/librahu/HIN-Datasets-for-Recommendation-and-Network-Embedding}{\color{blue}\textbf{Douban}}} & \multirow{2}{*}{2,848} & \multirow{2}{*}{39,586} & \multirow{2}{*}{894,887} & \multirow{2}{*}{35,770} & 
    \cite{bai2020tgnn,walker2021soapvae,seng2021atgcn,salamat2021heterographrec,xu2020srhgnn,gu2021egfrec,niu2021mgsr,han2022dhhgcn,sun2022motifres,liu2022hosr,liu2022gnnrec,liu2022siga,song2019dgrec,chen2022igrec}, \\ 
    & & & & & & \cite{yu2022esrf,yu2021sept,yu2021mhcn,song2020dream,miao2022melgn} \\
    \bottomrule
    \end{tabular}
    \begin{tablenotes}
        \item \textsuperscript{1}Raw dataset is available at {\color{blue} \url{https://www.yelp.com/dataset/documentation/main}}.
    \end{tablenotes}
    \end{threeparttable}
    }
\end{table*}

\section{Experimental Setup}\label{sec:setup}
In this section, we discuss the experimental setup of GNN-based \sorec methods.
Specifically, we review 17 benchmark datasets and 8 evaluation metrics, that are widely used in GNN-based \sorec methods.
Furthermore, we compare the recommendation accuracy between GNN-based SocialRS methods across datasets.

\subsection{\textbf{Benchmark Datasets}}

We summarize the datasets widely used by existing GNN-based \sorec methods in Table~\ref{tab:datasets}.
These datasets come from 8 different application domains: product, location, movie, image, music, bookmark, microblog, and miscellaneous.
We present the statistics of each dataset, including the numbers of users, items, ratings, and social relations, and a list of papers using the corresponding dataset. 
Since several versions exist per dataset, we chose the version that includes the most significant number of rating information. 

\subsubsection{\textbf{Product-related Datasets}}
\vspace{1mm}
\hfill \break
\indent\textbf{Epinions.}
This dataset is collected from a now-defunct consumer review site, Epinions.
It contains 355.8K trust relations from 18.0K users and 764.3K ratings from 18.0K users on 261.6K products. 
Here, a trust relation between two users indicates that one user trusts a review of a product written by another user.
For each rating, this dataset originally provides the product name, its category, the rating score in the range [1, 5], the timestamp that a user rated on an item, and the helpfulness of this rating.
37 GNN-based \sorec methods reviewed in this survey used this dataset~\cite{zheng2021pdarec,guo2020gnnsor,xiao2021mutualrec,fan2019graphrec,fu2021dicer,mandal2021gnntsr,mu2019gatnsr,bai2020tgnn,hou2021pagan,xiao2020mgnn,bi2021ghscf,li2020hidm,walker2021soapvae,fan2022graphrecp,hoang2021gtn,wu2020diffnet++, wu2019danser,tien2020kconvgraphrec,salamat2021heterographrec,xu2020srhgnn, zhao2021bfhan,narang2021fuserec,sun2020dgarecr,xiao2022gsfr,zhen2022apte,liu2022fesog,du2022sdcrec,liu2022sohrml,lin2022gnndsr,chen2022gdsrec,li2021spex,song2020dream,sha2021dsr,tao2022design,wang2021hypersorec,wang2022mohcn,liu2022gnnrec}, which means the most popular in \sorec.

\vspace{1mm}
\textbf{Ciao.}
This dataset is collected from a consumer review site in the UK, Ciao (\url{https://www.ciao.co.uk/}).
It contains 111.7K trust relations from 7.3K users and 283.3K ratings from 7.3K users on 104.9K products. The rating scale is from 1 to 5.
This dataset was used from 34 GNN-based \sorec methods reviewed in this survey~\cite{mandal2021gnntsr,bai2020tgnn,hou2021pagan,bi2021ghscf,li2020hidm,liao2022sociallgn,walker2021soapvae,liu2022mpsr,tien2020kconvgraphrec,fan2022graphrecp,salamat2021heterographrec,xu2020srhgnn,zhao2021bfhan,narang2021fuserec,sun2020dgarecr,wang2021hypersorec,fan2019graphrec,fu2021dicer,jiang2021asr,hoang2021gtn,wu2022dcrec,liu2022siga,xiao2022gsfr,wu2022eagcn,liu2022fesog,du2022sdcrec,liu2022sohrml,sha2021dsr,tao2022design,chen2022gdsrec,wang2022mohcn,song2022mrapr,lin2022gnndsr}.

\vspace{1mm}
\textbf{Beidan.}
This dataset is collected from a social e-commerce platform in China, Beidan (\url{https://www.beidian.com/}), which allows users' sharing behaviors. 
It includes 2.3K social relations from 2.8K users and 35.1K ratings from 2.8K users on 2.2K products.
For each social relation, Li et al~\cite{li2022disgcn} collected when a user's friend clicks a link shared by the user that points to the information of a specific item.
In this dataset, rating information does not provide explicit preference scores of users, rather containing implicit feedback only.
This dataset was used in only one GNN-based \sorec method~\cite{li2022disgcn}.

\vspace{1mm}
\textbf{Beibei.}
This dataset is collected from another social e-commerce platform in China, Beibei (\url{https://www.beibei.com/}).
It is similar to Beidan but provides larger sizes of social relations and ratings.
This dataset includes 197.5K social relations from 24.8K users and 1.6M ratings from 24.8K users on 16.8K products.
For ratings, this dataset provides users' implicit feedback.
This dataset was used in~\cite{li2022disgcn} only.

\subsubsection{\textbf{Location-related Datasets}}
\vspace{1mm}
\hfill \break
\indent\textbf{Yelp.}
This dataset is collected from a business review site, Yelp (\url{https://www.yelp.com/}).
It contains 363.6K social relations from 19.5K users and 405.8K ratings from 19.5K users on 21.2K businesses.
On Yelp, users can share their check-ins about local businesses (\eg restaurants and home services) and express their experience through ratings in the range [0, 5].
Also, users can create social relations with other users. 
Each check-in contains a user, a timestamp, and a business (\ie an item) that the user visited. 
29 GNN-based \sorec methods reviewed in this survey used this dataset~\cite{jiang2021asr,vijaikumar2019sorecgat,liu2022mpsr,guo2020gnnsor,wu2019diffnet,wu2020diffnet++,song2021diffnetlg,jin2020megcn,jiang2021san,yu2021sept,yu2021mhcn,han2022dhhgcn,sun2022motifres,zhen2022apte,wu2022eagcn,liu2022hosr,tao2022design,wang2021hypersorec,xie2022sran,li2022idiffnet,zhang2022cgl,wang2022mohcn,song2022mrapr,wei2022sghan,miao2022melgn,song2019dgrec,shi2022sengr,chen2022fbne,qiao2022tag}. 

\vspace{1mm}
\textbf{Dianping.}
This dataset is collected from a local restaurant search and review platform in China, Dianping (\url{https://www.dianping.com/}).
It contains 813.3K social relations from 59.4K users and 934.3K ratings from 59.4K users on 10.2K restaurants.
For ratings, each user can give scores in the range [1, 5].
This dataset was used in two GNN-based \sorec methods~\cite{wu2020diffnet++,wu2022dcrec}.

\vspace{1mm}
\textbf{Gowalla.}
This dataset is collected from a location-based social networking site, Gowalla (\url{https://www.gowalla.com/}).
It contains 283.7K friendship relations from 33.6K users and 1.2M ratings from 33.6K users on 41.2K locations. 
On Gowalla, users can share information about their locations by check-in and make friends based on the shared information. 
For ratings, this dataset provides users' implicit feedback.
9 GNN-based \sorec methods used this dataset~\cite{seng2021atgcn,chen2021serec,li2022atstggnn,wu2022eagcn,yu2022esrf,wang2022dcan,wei2022sghan,chen2022ssrgnn,liufu2021sga}.

\vspace{1mm}
\textbf{Foursquare.}
This dataset is collected from another location-based social networking site, Foursquare (\url{https://foursquare.com/}).
It is similar to Gowalla but provides larger sizes of social relations and ratings.
It contains 304.0K friendship relations from 39.3K users and 3.6M ratings from 39.3K users on 45.5K locations.
For ratings, this dataset provides users' implicit feedback.
This dataset was used in 4 GNN-based \sorec methods~\cite{chen2021serec,li2022atstggnn,wang2022dcan,chen2022ssrgnn}.

\subsubsection{\textbf{Movie-related Datasets}}
\vspace{1mm}
\hfill \break
\indent\textbf{MovieLens.}
This dataset is collected from GroupLens Research (\url{https://grouplens.org/}) for the purpose of recommendation research.
It contains 487.1K social relations from 138.1K users and 1.5M ratings from 138.1K users on 16.9K movies.
It should be noted that this dataset has different versions according to the size of the rating information. For the details, refer to \url{https://grouplens.org/datasets/movielens/}.
Since the original MovieLens datasets do not contain users' social relations, methods using this dataset built social relations by calculating the similarities between users.
This dataset was used in 4 GNN-based \sorec methods~\cite{liu2021sagclg,tien2020kconvgraphrec,jiang2022socialripplenet,chen2022fbne}.

\vspace{1mm}
\textbf{Flixster.}
This dataset is collected from a movie review site, Flixster (\url{https://www.flixster.com/}).
It contains 667.3K friendship relations from 58.4K users and 3.6M ratings from 58.4K users on 38.0K movies.
On Flixster, users can add other users to their friend lists and express their preferences for movies.
The rating values are 10 discrete numbers in the range [0.5, 5].
We found that 7 GNN-based \sorec methods used this dataset~\cite{xiao2020mgnn,fan2022graphrecp,guo2020gnnsor,xiao2021mutualrec,xiao2022gsfr,liu2022sohrml,liu2022siga}.

\vspace{1mm}
\textbf{FilmTrust.}
This dataset is collected from another (now-defunct) movie review site, FilmTrust.
It is similar to Flixster but provides smaller sizes of social relations and ratings.
It contains 1.8K friendship relations from 1.5K users and 35.4K ratings from 1.5K users on 2.0K movies.
The rating scale is from 1 to 5.
This dataset was used in 6 GNN-based \sorec methods~\cite{jiang2021asr,mu2019gatnsr,zheng2021pdarec,sun2022motifres,liu2022fesog,liu2022siga}.

\subsubsection{\textbf{Image-related Dataset}}
\vspace{1mm}
\hfill \break
\indent\textbf{Flickr.}
This dataset is collected from a who-trust-whom online image-based social sharing platform, Flickr (\url{https://www.flickr.com/}). 
It contains 187.2K follow relations from 8.3K users and 327.8K ratings from 8.3K users on 82.1K images.
On Flickr, users can follow other users and share their preferences for images with their followers.
For ratings, this dataset provides users' implicit feedback.
Also, we found that 9 GNN-based \sorec methods used this dataset~\cite{wu2019diffnet,wu2020diffnet++,jin2020megcn,jiang2021san,wu2022eagcn,tao2022design,xie2022sran,li2022idiffnet,zhang2022cgl}.

\subsubsection{\textbf{Music-related Dataset}}
\vspace{1mm}
\hfill \break
\indent\textbf{Last.fm.}
This dataset is collected from a social music platform, Lat.fm (\url{https://www.last.fm/}).
It contains 25.4K social relations from 1.8K users and 92.8K ratings from 1.8K users on 17.6K music artists.
Each rating indicates that one user listened to an artist's music, \ie implicit feedback.
On Lat.fm, users can make friend relations based on their preferences for artists.
This dataset was used in 11 GNN-based \sorec methods~\cite{liao2022sociallgn, xiao2021mutualrec, seng2021atgcn,tien2020kconvgraphrec,yu2021sept,yu2021mhcn,zhang2021kcrec,chen2022igrec,yu2022esrf,liufu2021sga,miao2022melgn}.

\subsubsection{\textbf{Bookmark-related Dataset}}
\vspace{1mm}
\hfill \break
\indent\textbf{Delicious.}
This dataset is collected from a social bookmarking system, Delicious (\url{https://del.icio.us/}). 
It contains 12.5K social relations from 1.6K users and 282.4K ratings from 1.6K users on 3.4K tags.
On Delicious, users can bookmark URLs (\ie implicit feedback) and also assign a variety of semantic tags to bookmarks.
Also, they can have social relations with other users having mutual bookmarks or tags.
This dataset was used in 7 GNN-based \sorec methods~\cite{li2020hidm,gu2021egfrec, chen2021serec,wang2022dcan,chen2022ssrgnn,lin2022gnndsr,song2019dgrec}.

\subsubsection{\textbf{Microblog-related Datasets}}
\vspace{1mm}
\hfill \break
\indent\textbf{Weibo.}
This dataset is collected from a social microblog site in China, Weibo (\url{https://weibo.com/}).
It contains 133.7K social relations from 6.8K users and 157.5K ratings from 6.8K users on 19.5K blogs.
On Weibo, users can post microblogs (\ie implicit feedback) and retweet other users' blogs. 
Based on such retweeting behavior, Li et al.~\cite{li2021spex} collected social relations between users.
Specifically, if a user has retweeted a microblog from another user, a social relation between the two users is created. 
This dataset was used in only one GNN-based \sorec method~\cite{li2021spex}.

\vspace{1mm}
\textbf{Twitter.}
This dataset is collected from another social microblog site, Twitter (\url{https://twitter.com/}).
It is similar to Weibo and contains 96.7K social relations from 8.3K users and 466.2K ratings from 8.9K users on 232.8K blogs.
Li et al.~\cite{li2021spex} collected social relations between two users if a user retweets or replies to a tweet from another user.
This dataset was used in~\cite{li2021spex} only.

\subsubsection{\textbf{Miscellaneous}}
\vspace{1mm}
\hfill \break
\indent\textbf{Douban.}
This dataset is collected from a social platform in China, Douban (\url{https://douban.com/}).
It contains 35.7K social relations from 2.8K users and 894.8K ratings from 2.8K users on 39.5K items of different categories (\eg books, movies, movies, and so on).
For ratings, this dataset provides users' implicit feedback.
This dataset was used in 19 GNN-based \sorec methods~\cite{bai2020tgnn,walker2021soapvae,seng2021atgcn,salamat2021heterographrec,xu2020srhgnn,gu2021egfrec,niu2021mgsr,han2022dhhgcn,sun2022motifres,liu2022hosr,liu2022gnnrec,liu2022siga,song2019dgrec,chen2022igrec,yu2022esrf,yu2021sept,yu2021mhcn,song2020dream,miao2022melgn}.
However, it should be noted that most methods using this dataset split users' ratings according to the item categories and then use those of some categories only, \eg Douban-Movie and Douban-Book.

\subsection{Evaluation Metrics}

\subsubsection{\textbf{Rating Prediction Task}} The methods that focus on explicit feedback aim to minimize the errors of the rating prediction task. 
To evaluate the performance of this task, they use the following metrics: root mean squared error (RMSE) and mean absolute error (MAE).
Specifically, MAE calculates the average error, the difference between the predicted and actual ratings, while RMSE emphasizes larger errors.
Both metrics are computed as follows:
\begin{equation}\label{maermse}
\begin{aligned}
MAE &= \frac{1}{M}\sum_{p_i\in\mathcal{U}}\sum_{q_j\in\mathcal{N}_{p_i}}|\hat{r}_{ij}-{r}_{ij}|,\\
RMSE &= \sqrt{\frac{1}{M}\sum_{p_i\in\mathcal{U}}\sum_{q_j\in\mathcal{N}_{p_i}}(\hat{r}_{ij}-{r}_{ij})^2},
\end{aligned}
\end{equation}
where $M$ indicates the number of ratings. Also, $\mathcal{U}$ and $\mathcal{N}_{p_i}$ denote a set of users and a set of items rated by $p_i$, respectively.
Lastly, ${r}_{ij}$ and $\hat{r}_{ij}$ indicate a user $p_i$'s actual and predicted ratings on an item $q_j$, respectively.

\subsubsection{\textbf{Top-$N$ Recommendation Task}} The methods for implicit feedback aim to improve the accuracy of the top-$N$ recommendation task. 
To evaluate the performance of this task, they use the following metrics: normalized discounted cumulative gain (NDCG)~\cite{JarvelinK00}, mean reciprocal rank (MRR)~\cite{BreeseHK98}, area under the ROC curve (AUC), F1 score, precision, recall, and hit rate (HR).

First, NDCG reflects the importance of ranked positions of items in a set  $\mathcal{R}_{p_i}$ of $N$ items that each method recommends to a user $p_i$. 
Let $y_{k}$ represent a binary variable for $k$-th item $i_{k}$ in $\mathcal{R}_{p_i}$, \ie $y_{k}\in{\{0,1\}}$. 
$y_{k}$ is set as $1$ if $i_{k}\in \mathcal{R}_{p_i}$ and set as $0$ otherwise. 
$\mathcal{N}_{p_i}$ denotes a set of items considered relevant to $p_i$ (\ie ground truth). 
In this case, $\text{NDCG}_{p_i}@N$ is computed by:
\begin{equation}\label{NDCG}
\begin{aligned}
\text{NDCG}_{p_i}@N & = \dfrac {\text{DCG}_{p_i}@N}{\text{IDCG}_{p_i}@N}, \\
\text{DCG}_{p_i}@N  & = \sum_{k=1}^{N}\dfrac {2^{y_{k}}-1}{\log_{2}{(k+1)}}, 
\end{aligned}
\end{equation}
where $\text{IDCG}_{p_i}@N$ is the ideal DCG at \textit{N}, \ie for the top-N items $i_{k} \in \mathcal{N}_{p_i}$, $y_{k}$ is set as $1$.

Second, MRR reflects the average inversed rankings of the first relevant item $i_{k}$ in $\mathcal{R}_{p_i}$. $\text{MRR}_{p_i}@N$ is computed by:
\begin{equation}\label{MRR}
\text{MRR}_{p_i}@N  = \dfrac {1}{\text{rank}_{p_i}},
\end{equation}
where $\text{rank}_{p_i}$ refers to the rank position of the first relevant item in $\mathcal{R}_{p_i}$. 

Third, AUC evaluates whether each method ranks a rated item higher than an unrated item. 
That is, AUC$_{p_i}$ is computed by:
\begin{equation}\label{auc}
\text{AUC}_{p_i}  = \dfrac {\sum_{q_j\in\mathcal{N}_{p_i}}\sum_{q_k\in\mathcal{N}_{p_i}\text{\textbackslash}{\mathcal{I}}}I(\hat{r}_{ij}>\hat{r}_{ik})}{\arrowvert \mathcal{N}_{p_i} \arrowvert \arrowvert \mathcal{N}_{p_i}\text{\textbackslash}{\mathcal{I}} \arrowvert},
\end{equation}
where $I(\cdot)$ is the indicator function. 

F1 score measures a harmonic mean of the precision and recall of the predictions as: 
\begin{equation}
\text{F1}_{p_i}@N = 2 \cdot \dfrac {\text{Precision}_{p_i}@N \cdot \text{Recall}_{p_i}@N}{\text{Precision}_{p_i}@N+\text{Recall}_{p_i}@N},
\end{equation}
\begin{equation}\label{precision}
\begin{aligned}
\text{Precision}_{p_i}@N & = \dfrac {\arrowvert \mathcal{N}_{p_i}\bigcap \mathcal{R}_{p_i} \arrowvert}{\arrowvert \mathcal{R}_{p_i} \arrowvert}, \\
\text{Recall}_{p_i}@N & = \dfrac {\arrowvert \mathcal{N}_{p_i}\bigcap \mathcal{R}_{p_i} \arrowvert}{\arrowvert \mathcal{N}_{p_i} \arrowvert},
\end{aligned}
\end{equation}
where $\text{Precision}_{p_i}@N$ and $\text{Recall}_{p_i}@N$ denote precision and recall at $N$, respectively.

Finally, HR is simply the fraction of users for which the ground truth is included in each $\mathcal{R}_{p_i}$:
\begin{equation}
    \text{HR}@N = \frac{\sum_{p_i\in\mathcal{U}} hit_{p_i}}{m},
\end{equation}
where $m$ indicates the number of users.
Also, $hit_{p_i}$ is assigned 1 if $\mathcal{R}_{p_i}$ contains any of the ground truth of $p_i$, and 0 otherwise. 

\textit{Conclusion.} These metrics thus give complementary insights. While NDCG measures the rank-discounted cumulative gain of the recommendations relative to the ideal gain, MRR finds the predicted rank of the most relevant item for each user. Thus, MRR is focused on the rank of only the first relevant item but NDCG can incorporate the relevance of all items in the ranked list. AUC, F1, Precision, and Recall metrics are rank-free classification metrics that distinguish the prediction of a ranked and an unranked item. While precision measures the proportion of predicted items that are relevant, recall measures the proportion of relevant items that are predicted. Finally, the hit rate measures how many users got their ground truth items predicted in the ranked list.

\subsection{Experimental Results}

\begin{table}[t]
    \centering
    \caption{Comparison of recommendation accuracy for GNN-based SocialRS methods. It should be noted that only methods with the same settings on each dataset are listed.}
    \label{tab:accuracy}
    \vspace{-0.2cm}
    \resizebox{0.52\textwidth}{!}{
    \renewcommand{\arraystretch}{1.0}
    \renewcommand{\aboverulesep}{0.0pt}
    \renewcommand{\belowrulesep}{0.0pt}
    \renewcommand{\belowbottomsep}{1.722pt}
    \renewcommand{\abovetopsep}{2.798pt}
    \newcolumntype{H}{>{\setbox0=\hbox\bgroup}c<{\egroup}@{}}
    \begin{tabular}{c|c|c|c|cc}
        \multicolumn{6}{c}{\textbf{(a) Epinions}} \\
        \toprule
        \multicolumn{3}{c|}{\textbf{Encoders}} & \multirow{2}{*}{\textbf{Models}} & \multirow{2}{*}{\textbf{MAE}} & \multirow{2}{*}{\textbf{RMSE}} \\ \cmidrule{1-3}
        \textbf{User Social} & \textbf{User Interest} & \textbf{Item Encoder} &  & & \\
        \midrule 
        \multirow{16}{*}{GANN} & \multirow{12}{*}{GANN} & \multirow{12}{*}{GANN} &  GraphRec~\cite{fan2019graphrec} & 0.8168 & 1.0631 \\
        & & & DANSER~\cite{wu2019danser} & 0.7781 & 1.0268 \\
        & & & KConvGraph~\cite{tien2020kconvgraphrec} & 0.8057 & 1.0104 \\
        & & & SR-HGNN~\cite{xu2020srhgnn} & 0.7983 & 1.0326 \\ 
        & & & GAT-NSR~\cite{mu2019gatnsr} & 0.7780 & 1.0190 \\
        & & & PA-GAN~\cite{hou2021pagan} & 0.8112 & 1.0612 \\
        & & & HeteroGraphRec~\cite{salamat2021heterographrec} & 0.8104 & 1.0483 \\
        & & & GTN~\cite{hoang2021gtn} & 0.8436 & 1.0139 \\
        & & & GraphRec+\cite{fan2022graphrecp} & 0.8093 & 1.0576 \\
        & & & GSFR~\cite{xiao2022gsfr} & 0.8018 & 1.0501 \\ \cmidrule{3-6}
        & & GCN & BFHAN~\cite{zhao2021bfhan} & 0.8046 & 1.0403 \\ \cmidrule{2-6}
        & \multirow{2}{*}{GRNN} & \multirow{2}{*}{GRNN} & GNN-DSR~\cite{lin2022gnndsr} & 0.8016 & 1.0579 \\
        & & & DGARec-R~\cite{sun2020dgarecr} & 0.7818 & 1.0261 \\ \cmidrule{2-6}
        & - & - & GHSCF~\cite{bi2021ghscf} & 0.7968 & 0.9731 \\ \midrule
        \multirow{1}{*}{GCN} & \multirow{1}{*}{RNN} & \multirow{1}{*}{GANN} & MOHCN~\cite{wang2022mohcn} & 0.7905 & 1.0327 \\         
        \bottomrule
    \end{tabular}
    }
    \resizebox{0.82\textwidth}{!}{
    \renewcommand{\arraystretch}{1.0}
    \renewcommand{\aboverulesep}{0.0pt}
    \renewcommand{\belowrulesep}{0.0pt}
    \renewcommand{\belowbottomsep}{1.722pt}
    \renewcommand{\abovetopsep}{2.798pt}
    \newcolumntype{H}{>{\setbox0=\hbox\bgroup}c<{\egroup}@{}}
    \begin{tabular}{c|c|c|c|cccccc}
        \multicolumn{10}{c}{\textbf{(b) Yelp}} \\
        \toprule
        \multicolumn{3}{c|}{\textbf{Encoders}} & \multirow{2}{*}{\textbf{Models}} & \multirow{2}{*}{\textbf{HR@5}} & \multirow{2}{*}{\textbf{HR@10}} & \multirow{2}{*}{\textbf{HR@15}} & \multirow{2}{*}{\textbf{NDCG@5}} & \multirow{2}{*}{\textbf{NDCG@10}} & \multirow{2}{*}{\textbf{NDCG@15}} \\ \cmidrule{1-3}
        \textbf{User Social} & \textbf{User Interest} & \textbf{Item Encoder} &  & & \\
        \midrule 
        \multirow{5}{*}{GANN} & \multirow{5}{*}{GANN} & \multirow{3}{*}{GANN} & 
        DiffNet++~\cite{wu2020diffnet++} & 0.2602&  0.3503 &	0.4051 &	0.1973	& 0.2288 &	0.2450  \\
        & & & DiffNetLG~\cite{song2021diffnetlg} & 0.2599 &	0.3711 &	0.4473 &	0.1941 &	0.2333	& 0.2586 \\
        & & & SRAN~\cite{xie2022sran} & 0.2639	& 0.3837 & 	0.4611 &	0.1953 &	0.2382 & 0.2614\\ \cmidrule{3-10}
        & & \multirow{2}{*}{Emb} & SAN~\cite{jiang2021san} & 0.2348 & 	0.3484 &	0.4257 &	0.1726 &	0.2125 &	0.2353 \\ 
        & & & MrAPR~\cite{song2022mrapr} & - & 0.3624 & - & - & 0.2871 & - \\  \midrule
        \multirow{2}{*}{GCN} & \multirow{2}{*}{GCN} & \multirow{2}{*}{Emb} & DiffNet~\cite{wu2019diffnet} & 0.2276 & 	0.3477 &	0.4232 &	0.1679 &	0.2121  &	0.2331 \\
        & & & MEGCN~\cite{jin2020megcn} & 0.2590 &	0.3685 &	0.4332 &	0.2012 &	0.2394	& 0.2590 \\ \midrule
        \multirow{1}{*}{LightGCN} & \multirow{1}{*}{LightGCN} & \multirow{1}{*}{Emb} & IDiffNet~\cite{li2022idiffnet} & 0.2625 &	0.3840 &	0.4640 &	0.1928 &	0.2360 &	0.2604 \\
        \bottomrule
    \end{tabular}
    }
    \resizebox{0.52\textwidth}{!}{
    \renewcommand{\arraystretch}{1.0}
    \renewcommand{\aboverulesep}{0.0pt}
    \renewcommand{\belowrulesep}{0.0pt}
    \renewcommand{\belowbottomsep}{1.722pt}
    \renewcommand{\abovetopsep}{2.798pt}
    \newcolumntype{H}{>{\setbox0=\hbox\bgroup}c<{\egroup}@{}}
    \begin{tabular}{c|c|c|c|cc}
        \multicolumn{6}{c}{\textbf{(c) Flixster}} \\
        \toprule
        \multicolumn{3}{c|}{\textbf{Encoders}} & \multirow{2}{*}{\textbf{Models}} & \multirow{2}{*}{\textbf{MAE}} & \multirow{2}{*}{\textbf{RMSE}} \\ \cmidrule{1-3}
        \textbf{User Social} & \textbf{User Interest} & \textbf{Item Encoder} &  & & \\
        \midrule 
        \multirow{2}{*}{GANN} & \multirow{2}{*}{GANN} & \multirow{2}{*}{GANN} &  GraphRec+\cite{fan2022graphrecp} & 0.7047 & 0.9303 \\
        & & & GSFR~\cite{xiao2022gsfr} & 0.6871 & 0.9176 \\ \midrule
        \multirow{1}{*}{GCN} & \multirow{1}{*}{GCN} & \multirow{1}{*}{GCN} & GNN-SOR~\cite{guo2020gnnsor} & 0.865 & 0.8710 \\  \midrule       
        \multirow{1}{*}{GAE} & \multirow{1}{*}{GAE} & \multirow{1}{*}{GAE} & SIGA~\cite{liu2022siga} & - & 0.9050 \\
        \bottomrule
    \end{tabular}
    }
    \resizebox{0.82\textwidth}{!}{
    \renewcommand{\arraystretch}{1.0}
    \renewcommand{\aboverulesep}{0.0pt}
    \renewcommand{\belowrulesep}{0.0pt}
    \renewcommand{\belowbottomsep}{1.722pt}
    \renewcommand{\abovetopsep}{2.798pt}
    \newcolumntype{H}{>{\setbox0=\hbox\bgroup}c<{\egroup}@{}}
    \begin{tabular}{c|c|c|c|cccccc}
        \multicolumn{10}{c}{\textbf{(d) Flickr}} \\
        \toprule
        \multicolumn{3}{c|}{\textbf{Encoders}} & \multirow{2}{*}{\textbf{Models}} & \multirow{2}{*}{\textbf{HR@5}} & \multirow{2}{*}{\textbf{HR@10}} & \multirow{2}{*}{\textbf{HR@15}} & \multirow{2}{*}{\textbf{NDCG@5}} & \multirow{2}{*}{\textbf{NDCG@10}} & \multirow{2}{*}{\textbf{NDCG@15}} \\ \cmidrule{1-3}
        \textbf{User Social} & \textbf{User Interest} & \textbf{Item Encoder} &  & & \\
        \midrule 
        \multirow{3}{*}{GANN} & \multirow{3}{*}{GANN} & \multirow{2}{*}{GANN} & 
        DiffNet++~\cite{wu2020diffnet++} & 0.1412 &   0.1832 &	0.2203 &	0.1269	& 0.1420 &	0.1544  \\
        & & & SRAN~\cite{xie2022sran} & 0.1540	& 0.1970 & 	0.2329 &	0.1395 &	0.1539 & 0.1653 \\ \cmidrule{3-10}
        & & \multirow{1}{*}{Emb} & SAN~\cite{jiang2021san} & 0.1267 & 	0.1653 &	0.1977 &	0.1151 &	0.1290 &	0.1393 \\ \midrule
        \multirow{2}{*}{GCN} & \multirow{2}{*}{GCN} & \multirow{2}{*}{Emb} & DiffNet~\cite{wu2019diffnet} & 0.1210 & 	0.1641 &	0.1952 &	0.1142 &	0.1273  &	0.1384 \\
        & & & MEGCN~\cite{jin2020megcn} & 0.1302 &	0.1688 &	0.2053 &	0.1208 &	0.1344	& 0.1460 \\ \midrule
        \multirow{2}{*}{LightGCN} & \multirow{2}{*}{LightGCN} & \multirow{1}{*}{LightGCN} & 
        DESIGN~\cite{tao2022design} & - & 0.2517 & - & - & 0.2590 & - \\ \cmidrule{3-10}
        & & Emb & IDiffNet~\cite{li2022idiffnet} & 0.1706 &	0.2269 &	0.2684 &	0.1499 &	0.1694 &	0.1833 \\
        \bottomrule
    \end{tabular}
    }
        \resizebox{0.92\textwidth}{!}{
    \renewcommand{\arraystretch}{1.0}
    \renewcommand{\aboverulesep}{0.0pt}
    \renewcommand{\belowrulesep}{0.0pt}
    \renewcommand{\belowbottomsep}{1.722pt}
    \renewcommand{\abovetopsep}{2.798pt}
    \newcolumntype{H}{>{\setbox0=\hbox\bgroup}c<{\egroup}@{}}
    \begin{tabular}{c|c|c|c|cccccc}
        \multicolumn{10}{c}{\textbf{(e) Last.fm}} \\
        \toprule
        \multicolumn{3}{c|}{\textbf{Encoders}} & \multirow{2}{*}{\textbf{Models}} & \multirow{2}{*}{\textbf{Precision@10}} & \multirow{2}{*}{\textbf{Precision@20}} & \multirow{2}{*}{\textbf{Recall@10}} & \multirow{2}{*}{\textbf{Recall@20}} & \multirow{2}{*}{\textbf{NDCG@10}} & \multirow{2}{*}{\textbf{NDCG@20}} \\ \cmidrule{1-3}
        \textbf{User Social} & \textbf{User Interest} & \textbf{Item Encoder} &  & & \\
        \midrule 
        \multirow{1}{*}{GANN} & \multirow{1}{*}{GANN} & \multirow{1}{*}{GANN} & 
        SGA~\cite{liufu2021sga} & - &   0.0712 & - &	0.2497	& - &	0.2723 \\ \midrule
        \multirow{1}{*}{GCN} & \multirow{1}{*}{GCN} & \multirow{1}{*}{Emb} & ATGCN~\cite{seng2021atgcn} & 0.1253 & 	- &	- &	- &	0.1607  &	- \\ \midrule
        \multirow{2}{*}{LightGCN} & \multirow{2}{*}{LightGCN} & \multirow{2}{*}{LightGCN} & 
        SEPT~\cite{yu2021sept} & 0.2019 & - & 0.2048 & - & 0.2450 & - \\
        & & & SocialLGN~\cite{liao2022sociallgn} & 0.1972 & 0.1368 & 0.2026 & 0.2794 & 0.2566 & 0.2883 \\ \midrule
        HyperGNN & GANN & HyperGNN & MHCN~\cite{yu2021mhcn} & 0.2005 &	- &	0.2037 &	- &	0.2439 &	- \\
        \bottomrule
    \end{tabular}
    }
            \resizebox{\textwidth}{!}{
    \renewcommand{\arraystretch}{1.0}
    \renewcommand{\aboverulesep}{0.0pt}
    \renewcommand{\belowrulesep}{0.0pt}
    \renewcommand{\belowbottomsep}{1.722pt}
    \renewcommand{\abovetopsep}{2.798pt}
    \newcolumntype{H}{>{\setbox0=\hbox\bgroup}c<{\egroup}@{}}
    \begin{tabular}{c|c|c|c|ccccccc}
        \multicolumn{11}{c}{\textbf{(f) Delicious}} \\
        \toprule
        \multicolumn{3}{c|}{\textbf{Encoders}} & \multirow{2}{*}{\textbf{Models}} & \multirow{2}{*}{\textbf{Recall@10}} & \multirow{2}{*}{\textbf{Recall@20}} & \multirow{2}{*}{\textbf{Recall@50}} & \multirow{2}{*}{\textbf{NDCG@10}} & \multirow{2}{*}{\textbf{NDCG@20}} & \multirow{2}{*}{\textbf{NDCG@50}} & \multirow{2}{*}{\textbf{MRR@20}} \\ \cmidrule{1-3}
        \textbf{User Social} & \textbf{User Interest} & \textbf{Item Encoder} &  & & \\
        \midrule 
        \multirow{2}{*}{GANN} & \multirow{1}{*}{GANN} & \multirow{1}{*}{Emb} & 
        HIDM~\cite{li2020hidm} & 0.1730 &   0.2255 & 0.3025 &	0.1145	& 0.1290 &	0.1453 & - \\ \cmidrule{2-11} 
        & GRNN & GRNN & GNN-DSR~\cite{lin2022gnndsr} & - & - & - & 0.2805 & 0.3164 & - & 0.2254 \\ \midrule
        
        \multirow{1}{*}{GCN} & \multirow{1}{*}{GRNN} & \multirow{1}{*}{GCN} & EGFRec~\cite{gu2021egfrec} & - & 0.4181	&	- &	- &	0.3000  &	- & 0.1573 \\ \midrule
        HetGNN & HetGNN & GANN & DCAN~\cite{wang2022dcan} & - & - & - & 0.2805 & 0.3045 & - & 0.2198 \\
        \bottomrule
    \end{tabular}
    }
    \resizebox{0.62\textwidth}{!}{
    \renewcommand{\arraystretch}{1.0}
    \renewcommand{\aboverulesep}{0.0pt}
    \renewcommand{\belowrulesep}{0.0pt}
    \renewcommand{\belowbottomsep}{1.722pt}
    \renewcommand{\abovetopsep}{2.798pt}
    \newcolumntype{H}{>{\setbox0=\hbox\bgroup}c<{\egroup}@{}}
    \begin{tabular}{c|c|c|c|ccc}
        \multicolumn{7}{c}{\textbf{(g) Douban}} \\
        \toprule
        \multicolumn{3}{c|}{\textbf{Encoders}} & \multirow{2}{*}{\textbf{Models}} & \multirow{2}{*}{\textbf{Precision@10}} & \multirow{2}{*}{\textbf{Recall@10}} & \multirow{2}{*}{\textbf{NDCG@10}} \\ \cmidrule{1-3}
        \textbf{User Social} & \textbf{User Interest} & \textbf{Item Encoder} &  & & \\
        \midrule 
        \multirow{3}{*}{GANN} & \multirow{3}{*}{GANN} & \multirow{3}{*}{GANN} &  TGRec~\cite{bai2020tgnn} & - & - & 0.2793 \\
        & & & ESRF~\cite{yu2022esrf} & 0.1823 & 0.0654 & 0.2103 \\ 
        & & & IGRec~\cite{chen2022igrec} & - & 0.4806 & 0.3921 \\       \midrule
        \multirow{3}{*}{GCN} & \multirow{1}{*}{GCN} & \multirow{1}{*}{GCN} & ATGCN~\cite{seng2021atgcn} & 0.1918 & - & 0.2084 \\ \cmidrule{2-7}
        & \multirow{2}{*}{GRNN} & GRNN & GNNRec~\cite{liu2022gnnrec} & - & 0.2350 & - \\ \cmidrule{3-7}
        & & GCN & EGFRec~\cite{gu2021egfrec} & - & - & 0.2008 \\\midrule         
        \multirow{3}{*}{HyperGNN} & \multirow{3}{*}{GANN} & \multirow{3}{*}{HyperGNN} & MHCN~\cite{yu2021mhcn} & 0.1850 & 0.0668 & 0.2103 \\
        & & & Motif-Res~\cite{sun2022motifres} & 0.2118 & 0.0510 & 0.2413 \\
        & & & DH-HGCN~\cite{han2022dhhgcn}  & - & 0.1081 & 0.0891 \\        \bottomrule
    \end{tabular}
    }
\end{table}

It should be noted that GNN-based SocialRS methods used different sets of datasets for experimentation and had different experimental settings, such as training/test ratio, top-$k$ values, and metrics.
For a fair comparison, we selected one dataset for each domain and compared the accuracy values of methods with the same settings on that dataset.

Table~\ref{tab:accuracy} shows the results on Epinions (Product), Yelp (Location), Flixster (Movie), Flickr (Image), Last.fm (Music), Delicious (Bookmark), and Douban (Miscellaneous).
To summarize, we did not find evidence of GNN encoders being optimized for a specific domain. Therefore, the best performer will vary depending on the domain and metric.
However, it is worth noting that on the Douban dataset, many SocialRS methods are based on HyperGNN. This trend may be due to the fact that the Douban dataset contains different types of user behavior for different types of items.
By employing a HyperGNN encoder, the authors might attempt to accurately capture various motifs.
\section{Future Directions}\label{sec:directions}
In this section, we discuss the limitations of GNN-based \sorec methods and present several future research directions.  

\subsection{Graph Augmentation in GNN-based \sorec} 
An intrinsic challenge of GNN-based \sorec methods lies in the \textit{sparsity of the input data} (\ie user-item interactions and user-user relations). 
To mitigate this problem, some GNN-based \sorec methods~\cite{yu2021sept,wu2022dcrec,du2022sdcrec,wu2022dcrec,zhang2022cgl,li2022disgcn,wang2022dcan,sun2022motifres,yu2021mhcn} have explored more supervision signals from the input data so that such signals can be utilized as different views from the original graph structure. 
Although many graph augmentation techniques~\cite{ding2022augmentation,zhao2022augmentation} such as node/edge deletion and graph rewiring have been proposed recently in a machine learning area, existing GNN-based \sorec methods only focus on adding edges between two users or between a user and an item~\cite{yu2021sept,yu2021mhcn,wu2022dcrec,zhang2022cgl,li2022disgcn,wang2022dcan,du2022sdcrec,wu2022dcrec,sun2022motifres}.
Therefore, it is a promising direction to leverage extra self-supervision signals based on various augmentation techniques to learn user and item embeddings more efficiently and effectively. 

\subsection{Trustworthy GNN-based \sorec}
Existing GNN-based \sorec methods have focused on improving their \textit{accuracy} by only relying on users' past feedback.
However, it is worth mentioning that there are other important ``beyond accuracy'' metrics, which we call trustworthiness\footnote{``Trustworthy'' is defined in the Oxford Dictionary as follows: an object or a person that you can rely on to be good, honest, sincere, etc~\cite{zhang2022trustworthy}.}
according to~\cite{zhang2022trustworthy}. 
Motivated by the importance of such metrics, various trustworthy GNN architectures have been proposed to incorporate core aspects of trustworthiness, including robustness, explainability, privacy, and fairness, in the context of GNN encoders~\cite{zhang2022trustworthy}.
One GNN-based \sorec method is proposed in this direction to specifically address the privacy issue~\cite{liu2022fesog}.
In particular, Liu et al.~\cite{liu2022fesog} devised a framework that stores user privacy data only in local devices individually and analyzes them together via federated learning. 
Thus, developing trustworthy GNN-based \sorec is a wide open for research. 
For example, consider robustness: bad actors may want to target certain products to certain users in a \sorec setting; how robust would existing GNN-based \sorec be against such attackers is an unanswered question and opens opportunities to create accurate as well as robust models.

\subsection{Heterogeneity}
In real-world graphs, nodes and their interactions are often multi-typed.
Such graphs, which are called heterogeneous graphs, convey rich information such as heterogeneous attributes, meta-path structures, and temporal properties.
Although HetGNN encoders have recently attracted attention in many domains (\eg healthcare and cybersecurity)~\cite{wang2020heterogeneity}, there have been only a few attempts to leverage such heterogeneity in \sorec~\cite{chen2021serec,wang2022dcan}.
Therefore, designing a HetGNN-based \sorec method remains an open question for the future.

\subsection{Efficiency and Scalability}
Most real-world graphs are too large and also grow rapidly. 
However, most GNN-based \sorec methods are too complicated, thus facing difficulty scaling to such large-scale graphs.
Some works have attempted to make more scalable versions of models, including SocialLGN~\cite{liao2022sociallgn}, SEPT~\cite{yu2021sept}, and DcRec~\cite{wu2022dcrec}, have attempted to remove the non-linear activation function, feature transformation, and self-connection, whereas Tao et al.~\cite{tao2022design} leveraged the knowledge distillation (KD) technique into \sorec.
However, designing a highly scalable GNN architecture is an important problem that remains challenging to date. 

\section{Conclusions}\label{sec:conclusions}

Although there has been a surge of papers on developing GNN-based social recommendation methods, no survey paper existed that reviewed them thoroughly.  
Our work is the first systematic and comprehensive survey that studies $84$ papers on GNN-based \sorec, collected by following the PRISMA guidelines.
We present a novel taxonomy of inputs and architectures for GNN-based \sorec, thus, categorizing different methods developed over the years in this important topic. 
Through this survey, we hope to enable the researchers of this field to better position their works in the recent trend while forming a gateway for the new researchers to get introduced to this important and hot topic. 
We hope this survey helps readers to grasp recent trends in \sorec and develop novel GNN-based \sorec methods.

\begin{acks}
The work of Srijan Kumar is supported in part by NSF grants CNS-2154118, IIS-2027689, ITE-2137724, ITE-2230692, CNS-2239879, Defense Advanced Research Projects Agency (DARPA) under Agreement No. HR00112290102 (subcontract No. PO70745), and funding from Microsoft, Google, and The Home Depot.
The work of Sang-Wook Kim was supported by the Institute of Information \& communications Technology Planning \& Evaluation (IITP) grant funded by the Korean government (MSIT) (No.RS-2022-00155586, A High-Performance Big-Hypergraph Mining Platform for Real-World Downstream Tasks; No. 2020-0-01373, Artificial Intelligence Graduate School Program (Hanyang University)). The work of Yeon-Chang Lee was supported by the Institute of Information \& communications Technology Planning \& Evaluation (IITP) grant funded by the Korean government (MSIT)
(No.2020-0-01336, Artificial Intelligence Graduate School Program (UNIST)). 
\end{acks}

\bibliographystyle{ACM-Reference-Format}
\bibliography{citations}

\end{document}